\documentclass[10pt,nocopyrightspace,preprint]{sigplanconf}

\usepackage{amssymb,amsmath,amsthm}
\usepackage{mathtools} 
\usepackage{latexsym}
\usepackage{graphicx}
\usepackage[usenames,dvipsnames]{color}
\usepackage{listings}
\usepackage{float}
\usepackage{multirow}
\usepackage[scaled]{helvet}
\usepackage[noend]{algorithmic}
\usepackage{mathrsfs}
\usepackage{mathpartir}
\usepackage{dsfont} 
\usepackage{stmaryrd}
\usepackage{url}
\usepackage{textcomp} 
\usepackage[colorlinks=true,allcolors=blue,breaklinks,draft=false]{hyperref}
\usepackage{titlesec}
\usepackage{parskip}
\usepackage{alltt} 
\usepackage{bbm}
\usepackage{alltt}
\usepackage{verbdef}
\usepackage{xspace}
\usepackage{verbatim}
\usepackage{enumitem}
\usepackage{lipsum}
\usepackage{wrapfig}
\usepackage[usenames,dvipsnames]{xcolor}
\hypersetup{linkcolor=black,citecolor=black,urlcolor=black}
\newcounter{tags}

\usepackage{adjustbox} 
\usepackage{array}
\usepackage{booktabs}
\usepackage{multirow}
\usepackage{pifont}
 
\theoremstyle{plain} 
\newtheorem{theorem}{Theorem}[section]
\newtheorem{lemma}[theorem]{Lemma}

{\unskip\nobreak\hskip 1em plus 1fil\nobreak$\square$
\parfillskip=0pt%
\endtrivlist}

\newcommand{\code}[1]{\lstinline[basicstyle=\small\ttfamily]{#1}}

\newcommand{\etc}{\emph{etc}}
\newcommand{\ie}{\emph{i.e.}\xspace}

\newcommand{\eg}{\emph{e.g.}\xspace}

\newcommand{\etal}{\emph{et~al.}\xspace}

\newcommand{\wrt}{{wrt.}\xspace}

\newcommand{\res}{\mathsf{res}}
\newcommand{\bal}{\mathit{bal}}

\newcommand{\ic}{\mathcal{I}}
\newcommand{\Ic}[2]{\ic~{#1}~{#2}}
\newcommand{\hide}{\mathsf{hide}}  
\newcommand{\last}{\mathsf{last}}  

\newcommand{\specK}[1]{\ensuremath{\textcolor{blue}{#1}}}
\newcommand{\comm}[1]{\ensuremath{\textcolor{gray}{\esc{/\!/}~{#1}}}}
\newcommand{\spec}[1]{\specK{\left\{{#1}\right\}}}

\newcommand{\Num}[1]{{\text{{\scriptsize{#1}}}}}
\newcommand{\esc}[1]{\text{\texttt{\small{#1}}}}
\newcommand{\kw}[1]{\text{\textbf{#1}}}

\newcommand{\Asgn}{\leftarrow} 

\newcommand{\dotcup}{\ensuremath{\mathaccent\cdot\cup}}

\newcommand{\intab}[1]{({\sffamily{\small{#1}}})}

\newcommand{\rcon}{\mathcal{R}}

\theoremstyle{remark}
\newtheorem{example}{Example}[section]

\newcommand{\lcl}{{\mathsf{S}}}
\newcommand{\env}{{\mathsf{O}}}
\newcommand{\joint}{{\mathsf{J}}}
\newcommand{\hist}{\chi} 
\newcommand{\heap}{h} 
\newcommand{\perm}{\pi}
\newcommand{\gist}{\eta} 
\newcommand{\tkn}{\tau}
\newcommand{\ikn}{\iota}
\newcommand{\ikno}{\ikn_\env}
\newcommand{\iknh}{\hat{\iota}}
\newcommand{\tknh}{\hat{\iota}}

\newcommand{\Tomb}{\mathsf{spent}}

\newcommand{\histL}{\hist_\lcl}
\newcommand{\histE}{\hist_\env}
\newcommand{\hists}{\histL}
\newcommand{\gists}{\gist_\lcl}
\newcommand{\gisto}{\gist_\env}
\newcommand{\histo}{\histE}
\newcommand{\tkns}{\tkn_\lcl}
\newcommand{\tkno}{\tkn_\env}
\newcommand{\perms}{\perm_\lcl}
\newcommand{\permo}{\perm_\env}
\newcommand{\heaps}{\heap_\lcl}
\newcommand{\heapo}{\heap_\env}
\newcommand{\heapj}{\heap_\joint}

\makeatletter
\newcommand{\oset}[3][0ex]{%
  \mathrel{\mathop{#3}\limits^{
    \vbox to#1{\kern-3\ex@
    \hbox{$\scriptstyle#2$}\vss}}}}
\makeatother

\makeatletter
\newcommand{\ojset}[3][0ex]{%
  \mathrel{\mathop{#3}\limits^{
    \vbox to#1{\kern-5\ex@
    \hbox{$\scriptstyle#2$}\vss}}}}
\makeatother

\newcommand{\hunion}{\mathbin{\dotcup}}

\newcommand{\ldot}{\mathord{.}\,}
\newcommand{\aand}{,}
\newcommand{\oor}{\vee}

\newcommand{\eqdef}{\mathrel{\:\widehat{=}\:}}

\newcommand{\hpts}{\mapsto}
\newcommand{\set}[1]{\left\{#1\right\}}

\newcommand{\happrox}{\mathsf{ResPast}}

\newcommand{\eqqc}{e_{\mathit{qqc}}}
\newcommand{\eqc}{e_{\mathit{qc}}}

\newcommand{\pending}{{m_\joint}} 

\newcommand{\twin}[1]{\bar{#1}}
\newcommand{\mygather}[1]{|\!|{#1}|\!|}

\definecolor{shadecolor}{gray}{1.00}
\definecolor{ddarkgray}{gray}{0.75}
\definecolor{darkgray}{gray}{0.30}
\definecolor{light-gray}{gray}{0.87}

\newcommand{\graybox}[1]{\colorbox{light-gray}{#1}}

\newcommand{\gbm}[1]{\graybox{${#1}$}}

\setlist[itemize]{leftmargin=*}

\titlespacing*{\section}{0pt}{*1.5}{*1.5} 
\titlespacing*{\subsection}{3pt}{*0.8}{*0.5}
\titlespacing*{\subsubsection}{0pt}{*0.8}{*0.5}
\titlespacing*{\paragraph}{0pt}{*0.5}{*1.2}

\definecolor{shadecolor}{gray}{1.00}
\definecolor{darkgray}{gray}{0.30}
\definecolor{violet}{rgb}{0.56, 0.0, 1.0}
\definecolor{forestgreen}{rgb}{0.13, 0.55, 0.13}

\lstdefinelanguage{Coq} {
mathescape=true,						
texcl=false,
morekeywords=[1]{
  Add,
  All,
  Arguments,
  Axiom,
  Bind,
  Canonical,
  Check,
  Close,
  CoFixpoint,
  CoInductive,
  Coercion,
  Contextual,
  Corollary,
  Defined,
  Definition,
  Delimit,
  End,
  Example,
  Export,
  Fact,
  Fixpoint,
  Goal,
  Graph,
  Hint,
  Hypotheses,
  Hypothesis,
  Implicit,
  Implicits,
  Import,
  Inductive,
  Lemma,
  Let,
  Local,
  Locate,
  Ltac,
  Maximal
  Module,
  Morphism,
  Next,
  Notation,
  Obligation,
  Open,
  Parameter,
  Parameters,
  Prenex,
  Print,
  Printing,
  Program,
  Projections,
  Proof,
  Proposition,
  Qed,
  Record,
  Relation,
  Remark,
  Require,
  Reserved,
  Resolve,
  Rewrite,
  Save,
  Scope,
  Search,
  Section,
  Show,
  Strict,
  Structure,
  Tactic,
  Theorem,
  Unset,
  Variable,
  Variables,
  View,
  inside,
  outside
},
morekeywords=[2]{
  as,
  cofix,
  else,
  end,
  exists,
  exists2,
  fix,
  for,
  forall,
  fun,
  if,
  in,
  is,
  let,
  match,
  nosimpl,
  of,
  return,
  struct,
  then,
  vfun,
  with
},
morekeywords=[3]{Type, Prop, Set, True, False},
morekeywords=[4]{
  after,
  apply,
  assert,
  auto,
  bool_congr,
  case,
  change,
  clear,
  compute,
  congr,
  cut,
  cutrewrite,
  destruct,
  elim,
  field,
  fold,
  generalize,
  have,
  heval, 
  hnf,
  induction,
  injection,
  intro,
  intros,
  intuition,
  inversion,
  left,
  loss,
  move,
  nat_congr,
  nat_norm,
  pattern,
  pose,
  refine,
  rename,
  replace,
  revert,
  rewrite,
  right,
  ring,
  set,
  simpl,
  split,
  suff,
  suffices,
  symmetry,
  transitivity,
  trivial,
  unfold,
  unlock,
  using,
  without,
  wlog,
  autorewrite
},        
morekeywords=[5]{
  assumption,
  by,
  contradiction,
  done,
  exact,
  lia,
  gappa,
  omega,
  reflexivity,
  romega,
  solve,
  tauto,
  discriminate,
  unsat
},
morecomment=[s]{(*}{*)},
morekeywords=[6]{do, last, first, try, idtac, repeat},
showstringspaces=false,
morestring=[b]",
tabsize=3,							
extendedchars=true,  		 		
sensitive=true, 
breaklines=false,
basicstyle=\scriptsize\ttfamily,
captionpos=b,							
columns=[l]fullflexible,
identifierstyle={\color{black}},
keywordstyle=[1]{\color{violet}},
keywordstyle=[2]{\color{forestgreen}},
keywordstyle=[3]{\color{forestgreen}},
keywordstyle=[4]{\color{blue}},
keywordstyle=[5]{\color{red}},
keywordstyle=[6]{\color{violet}},
stringstyle=,
commentstyle=\it\ttfamily\color{Bittersweet},
numberstyle=\tiny,
}

\lstdefinestyle{Coq}{language=Coq}
\setcitestyle{square}
\defcitealias{Coq-manual}{Coq proof assistant}

\begin{document}

\authorinfo{Ilya Sergey$^\dag$ \and Aleksandar Nanevski$^\ddag$ \and Anindya Banerjee$^\ddag$
\and Germ\'{a}n Andr\'{e}s Delbianco$^\ddag$} 
{$^\dag$University College London, UK \and\and $^\ddag$IMDEA Software Institute, Spain}
{i.sergey@ucl.ac.uk \and \{aleks.nanevski, anindya.banerjee, german.delbianco\}@imdea.org}





\title{
  Hoare-style Specifications as Correctness Conditions \\
  for Non-linearizable Concurrent Objects }

\maketitle

\begin{abstract}

  Designing efficient concurrent objects often requires abandoning the
  standard specification technique of \emph{linearizability} in favor
  of more relaxed correctness conditions.
  However, the variety of alternatives makes it difficult to choose
  which condition to employ, and how to compose them when using
  objects specified by different conditions.


  In this work, we propose a \emph{uniform} alternative in the form of
  Hoare logic, which can explicitly capture---in the auxiliary
  state---the interference of environment threads.
  We demonstrate the expressiveness of our method by verifying a
  number of concurrent objects and their clients, which have so far
  been specified only by non-standard conditions of
  \emph{concurrency-aware linearizability}, \emph{quiescent}, and
  \emph{quantitative quiescent consistency}.
  We report on the implementation of the ideas in an existing
  Coq-based tool, providing the first mechanized proofs for all the
  examples in the paper.

\end{abstract}

\section{Introduction}
\label{sec:introduction}

Linearizability~\cite{Herlihy-Wing:TOPLAS90} remains the
most well-known correctness condition for concurrent objects.
It works by relating a concurrent object to a sequential behavior.
More precisely, for each concurrent history of an object,
linearizability requires that there exists a mapping to a sequential
history, such that the ordering of matching call/return pairs is
preserved either if they are performed by the same thread, or if they
do not overlap.
As such, linearizability has been used to establish the correctness of
a variety of concurrent objects such as stacks, queues, sets, locks,
and snapshots---all of which have intuitive sequential specs.

However, as argued by Shavit~\cite{Shavit:CACM11}, efficient
parallelization may require the development of concurrent objects
that are inherently \emph{non-linearizable}: in the presence of
interference, such objects exhibit behavior that is not reducible to
any sequential behavior via linearizability. To reason about such
objects, a variety of novel conditions has been developed:
concurrency-aware linearizability (CAL)~\cite{Hemed-Rinetzky:PODC14},
quiescent consistency (QC)~\cite{Aspnes-al:JACM94,Derrick-al:FM14},
quasi-linearizability (QL)~\cite{Afek-al:OPODIS10}, quantitative
relaxation~\cite{Henzinger-al:POPL13}, quantitative quiescent
consistency (QQC)~\cite{Jagadeesan-Riely:ICALP14}, and local
linearizability~\cite{Haas-al-local15}, to name a few.
These conditions, formulated as relations on execution traces, specify
a program's behavior under concurrent interference. Some, such as QC,
devote special treatment to the sequential case, qualifying the
behavior in the quiescent (\ie, interference-free) moments.
%

This proliferation of alternative conditions is problematic, as it
makes all of them non-canonical. For any specific example, it is
difficult to determine which condition to use, or if a new one should
be developed. Worse, each new condition requires a development of its
own dedicated program logic or verification tool. Furthermore, it is
unclear how to combine the conditions/logics/tools, when different
ones have been used for different subprograms.
Finally, having criteria defined \emph{semantically}, \eg, in terms of
execution traces, makes it challenging to employ them directly for
reasoning about clients of the corresponding data structures.

\subsection{Concurrency specification via program logics}

In this paper, we propose an alternative, uniform, approach: a Hoare
logic equipped with special \emph{subjective} kind of auxiliary
state~\cite{LeyWild-Nanevski:POPL13} that makes it possible to name
the amount of concurrent interference, and relate it to the program's
inputs and outputs \emph{directly}, without reducing to sequential
behavior. We use Fine-grained Concurrent Separation Logic
(FCSL)~\cite{Nanevski-al:ESOP14}, which has been designed to reason
about higher-order lock-free concurrent programs, and has been
recently implemented as a verification tool on top of
Coq~\cite{Sergey-al:PLDI15}, but whose ability to address
non-linearizable programs has not been observed previously.

More specifically, subjective auxiliary state permits that within a
spec of a thread, one can refer to the private state (real or
auxiliary) of \emph{other} interfering threads \emph{in a local
  manner}. This private state can have arbitrary user-specified
structure, as long as it satisfies the properties of a partial
commutative monoid (PCM).
A particularly important PCM is that of \emph{time-stamped histories},
which has previously been applied to linearizable
objects~\cite{Sergey-al:ESOP15}, where it replaced call/return
histories. A (logically) time-stamped history consists of entries of
the form $t\,{\mapsto}\,a$, signifying that an atomic behavior $a$
occurred at a time (or linearization point)~$t$. A subjective
specification further distinguishes the histories of the thread and
its interfering environment, and usefully relates both to the thread's
input and output.

Of course, Hoare-style reasoning about histories is a natural idea,
exploited recently in several
works~\cite{Fu-al:CONCUR10,Gotsman-al:ESOP13,Bell-al:SAS10,Hemed-al:DISC15}. Here,
however, we rely on the unifying power of PCMs, in combination with
subjective specifications, to show that by generalizing histories in
different ways---though all subject to PCM laws---we can capture the
essence of several different conditions, such as CAL, QC and QQC in
one-and-same \emph{off-the-shelf} logical system and tool.
More precisely, our histories need not merely identify a point at
which an atomic behavior logically occurred, but can also include
information about interference, or lack thereof. Moreover, we will use
generic FCSL constructs for delimiting the scope of auxiliary state,
to reason about quiescent moments.
%
%

\subsection{Contributions and outline}
\label{sec:chall-contr}

The ability to use FCSL for specifying and verifying
\emph{linearizable} objects (\eg, fine-grained stacks and atomic
snapshots) has been recognized before~\cite{Sergey-al:ESOP15}.
In contrast, the main conceptual contribution of this work is an
observation that the \emph{very same} abstractions provided by FCSL
are sufficient to ascribe non-trivial \emph{non-linearizable} objects
with specs that can hide object implementation details, but are
sufficiently strong to be used in proofs of concurrent client
programs, as we demonstrate in Section~\ref{sec:overview}.
Specifically, we recognize that auxiliary histories can be subject of
user-defined invariants beyond mere adherence to sequential executions
(\eg, be concurrency-aware~\cite{Hemed-Rinetzky:PODC14}), and can be
used to capture intermediate interference, allowing for quantitative
reasoning about outcomes of concurrent executions (\eg, in the spirit
of QQC~\cite{Jagadeesan-Riely:ICALP14}). These observations,
surprisingly, enabled reasoning about non-linearizable data structures
and their clients, which were never previously approached from the
perspective of program logics or mechanically verified.

In this unified approach based on program logic, it seems inherently
impossible (and contrary to the whole idea) to classify Hoare triples
as corresponding to this or that correctness condition. Thus, instead
of providing theorems that relate Hoare triples to existing
conditions, we justify the adequacy of our approach by
proof-of-concept verifications of concurrent objects and their
clients.

Hence, as key technical contributions, we present \emph{subjective
  specs and the first mechanized proofs} (in Coq) of (1) an
elimination-based exchanger~\cite{Scherer-al:SCOOL05}
(Section~\ref{sec:exchanger}), previously specified using CAL, and (2)
a simple counting network~\cite{Aspnes-al:JACM94}
(Section~\ref{sec:counting}) that inspired definitions of QC and
QQC. We then employ these specs to verify client
programs~(Sections~\ref{sec:cal} and~\ref{sec:qclients}).
%
%
We discuss alternative design choices for specs and further
applications of our verification approach in
Section~\ref{sec:discussion}, and summarize our mechanization
experience in Section~\ref{sec:evaluation}. Section~\ref{sec:related}
compares to related work and Section~\ref{sec:conclusion} concludes.

\section{Main Ideas and Overview}
\label{sec:overview}

We begin by outlining the high-level intuition of our specification
approach, and summarize the main formalization steps.
As the first motivating example, we consider the concurrent exchanger
structure from $\mathtt{java.util.concurrent}$
\cite{Scherer-al:SCOOL05,ExchangerClass}. The main purpose of the
exchanger is to allow two threads to efficiently swap values in a
non-blocking way via a globally shared channel. The exchange might
fail, if a thread trying to swap a value does not encounter a peer to
do that in a predefined period of time.
%

For instance, the result of the two-thread program
\[
\tag{\arabic{tags}}\refstepcounter{tags}\label{tag:ex1} 
\begin{array}{c@{\ }c@{\ }c} 
  \boxed{T_1} & & \boxed{T_2}
  \\[5pt] 
   r_1 := \esc{exchange}~1 & || & r_2 := \esc{exchange}~2
\end{array}
\]
can be described by the following assertion:\footnote{We
  use ML-style \texttt{option} data type with two constructors,
  \texttt{Some} and \texttt{None} to indicate success and failure of
  an operation, correspondingly.}

\[
\tag{\arabic{tags}}\refstepcounter{tags}\label{tag:exsp} 
r_1 = r_2 = \esc{None} \oor r_1 = \esc{Some}~2 \wedge r_2 =
\esc{Some}~1
\]

\noindent
That is, $r_1$ and $r_2$ store the results of the execution of
subthreads $T_1$ and $T_2$ correspondingly, and both threads either
succeed, exchanging the values, or fail. The ascribed outcome is only
correct under the assumption that no other threads besides $T_1$ and
$T_2$ attempt to use the very same exchange channel concurrently.

Why is the exchanger not a linearizable data structure? To see that,
recall that linearizability reduces the concurrent behavior to a
sequential one~\cite{Herlihy-Wing:TOPLAS90}. If the exchanger were
linearizable, all possible outcomes of the program~\eqref{tag:ex1}
would be captured by the following two sequential programs, modelling
selected interleavings of the threads $T_1$ and $T_2$:
\[
\tag{\arabic{tags}}\refstepcounter{tags}\label{tag:ex2} 
\begin{array}{c}
r_1 := \esc{exchange}~1;~r_2 := \esc{exchange}~2;
\\[3pt]
\text{and}
\\[3pt]
r_2 := \esc{exchange}~2;~r_1 := \esc{exchange}~1;  
\end{array}
\]
However, both programs~\eqref{tag:ex2} will \emph{always} result in
$r_1 = r_2 = \esc{None}$, as, in order to succeed, a call to the
exchanger needs another thread, running concurrently, with which to
exchange values.
%
%
This observation demonstrates that linearizability with respect to a
sequential specification is too weak a correctness criterion to
capture the \code{exchanger}'s behavior observed in a truly concurrent
context~\cite{Hemed-Rinetzky:PODC14}: an adequate notion of
correctness for \code{exchange} must mention the effect
of {interference}.
%

Consider another structure, whose concurrent behavior cannot be
related to sequential executions via linearizability:
\[
\tag{\arabic{tags}}\refstepcounter{tags}\label{tag:flip} 
\begin{array}{l}
\esc{flip2}~(x : \esc{ptr~nat})~:~\esc{nat}~=~\{ \\[2pt]
~~ a := \esc{flip}~x; \\[2pt]
~~ b := \esc{flip}~x;\\[2pt]
~~\kw{return}~a + b ~~\}  
\end{array}
\]
The procedure \code{flip2} takes a pointer $x$, whose value is
either~0 or~1 and changes its value to the opposite, twice, via the
\emph{atomic} operation \code{flip}, returning the sum of the previous
values. Assuming that $x$ is being modified only by the calls to
\code{flip2}, what is the outcome $r$ of the following program?
\[
\tag{\arabic{tags}}\refstepcounter{tags}\label{tag:flip2} 
r := \esc{flip2}~x;
\]
The answer depends on the presence or absence of interfering threads
that invoke \code{flip2} concurrently with the
program~\eqref{tag:flip2}. Indeed, in the absence of interference,
\code{flip2} will flip the value of $x$ twice, returning the sum of 0
and 1, \ie,~1. However, in the presence of other threads calling
\code{flip2} in parallel, the value of $r$ may vary from 0 to 2.
%

What are the intrinsic properties of \code{flip2} to be specified?
Since the effect of \code{flip2} is distributed between two internal
calls to \code{flip}, both subject to interference, the specification
should capture that the variation in \code{flip2}'s result is
subject to interference.
%
%
%
%
Furthermore, the specification should be expressive enough to allow
reasoning under bounded interference. For example, absent interference
from any other threads besides $T_1$ and $T_2$ that invoke
\code{flip2} concurrently, the program below will always result in~$r
= 2$:
%
%
\[
\tag{\arabic{tags}}\refstepcounter{tags}\label{tag:flip3} 
\begin{array}{c@{\ }c@{\ }c@{\ }l@{\ }l} 
  \boxed{T_1} & & \boxed{T_2}
  \\[5pt] 
  r_1 := \esc{flip2}~x & || & r_2 := \esc{flip2}~x; \\[3pt]
  \multicolumn{3}{c}{r :=  r_1 + r_2}
\end{array}
\]

\subsection{Abstract histories of non-linearizable objects}
\label{sec:hist}

Execution histories capture the traces of a concurrent object's
interaction with various threads, and are a central notion for
specifying concurrent data structures.
For example, linearizability specifies the behavior of an object by
mapping the object's global history of method invocations and returns
to a
%
%
sequence of operations that can be observed when the object is used
sequentially~\cite{Herlihy-Wing:TOPLAS90}. However, as we have shown,
neither \code{exchange} nor \code{flip2} can be understood in terms of
sequential executions.

We propose to specify the behavior and outcome of such objects in
terms of \emph{abstract concurrent histories}, as follows. Instead of
tracking method invocations and returns, our histories track the
``interesting'' changes to the object's state. What is ``interesting''
is determined by the user, depending on the intended clients of
the concurrent object.
%
%
%
%
Moreover, our specifications are subjective (\ie, thread-relative) in
the following sense. Our histories do not identify threads by their
thread IDs. Instead, each method is specified by relating two
different history variables: the history of the invoking thread
(aka.~\emph{self}-history), and the history of its concurrent
environment (aka.~\emph{other}-history). In each thread, these two
variables have different values.
%

%
%

For example, in the case of the exchanger, the interesting changes to
the object's state are the exchanges themselves. Thus, the global
history $\hist_{\cal E}$ tracks the successful exchanges in the form
of pairs of values, as shown in below:
\[
\!\!\!\begin{array}{r@{\ }c@{}c@{\ }c@{\ }c@{\ }c@{\ }c@{\ }c@{\ }c@{}l}
  & & \boxed{T_1} & \graybox{\boxed{T_2}} & \graybox{\boxed{T_2}} & \graybox{\boxed{T_3}} & \graybox{\boxed{T_2}}
  & \boxed{T_1} 
  \\[5pt]  
  \hist_{\cal E} \!=\! \text{[} &\!\!...,~& (1, 2), & \graybox{(2, 1)}, & 
                                 \graybox{(4, 5)}, & \graybox{(5, 4)}, &  \graybox{(9, 8)}, & (8, 9), &
                                                                          ...
                     &
                       \text{]}   
\\[-5pt]   
&& \multicolumn{2}{l}{\underbracket{\phantom{aaaaaaaaa}}_{\mathtt{exchange~ok}}} & \multicolumn{2}{c}{\underbracket{\phantom{aaaaaaaaa}}_{\mathtt{exchange~ok}}} & \multicolumn{2}{c}{\underbracket{\phantom{aaaaaaaaa}}_{\mathtt{exchange~ok}}}
                  & & 
\end{array}
\]
The diagram presents the history from the viewpoint of thread
$T_1$. The exchanges made by $T_1$ are colored white, determining the
\emph{self}-history of $T_1$. The gray parts are the exchanges made by
the other threads (e.g., $T_2$, $T_3$, \etc.), and determine the
\emph{other}-history for $T_1$. 



The subjective division between \emph{self} and \emph{other} histories
emphasizes that a successful exchange is actually represented by
\emph{two} pairs of numbers $(x, y)$ and $(y, x)$, that appear
consecutively in $\hist_{\cal E}$, and encode the two ends of an
exchange from the viewpoint of the exchanging threads. We call such
pairs \emph{twins}.
%
As an illustration, the white entry $(2, 1)$ from the self-history of
$T_1$, is matched by a twin gray entry $(1, 2)$ from the other-history
of $T_1$, encoding that $T_1$ exchanging $2$ for $1$ corresponds to
$T_1$'s environment exchanging $1$ for $2$.

The subjective division is important, because it will enable us to
specify threads \emph{locally}, \ie, without referring to the code of
other threads. For example, in the case of program~(\ref{tag:ex1}), we
will specify that $T_1$, in the case of a successful exchange, adds a
pair $(1, r_1)$ to its self history, where $\esc{Some}~r_1$ is $T_1$'s
return value. Similarly, $T_2$ adds a pair $(2, r_2)$ to
its self history, where $\esc{Some}~r_2$ is $T_2$'s return value.
%

On the other hand, it is an important invariant of the exchanger
object---but not of any individual thread---that twin entries are
symmetric pairs encoding different viewpoints of the one-and-the-same
exchange. This object invariant will allow us to reason about clients
containing combinations of exchanging threads. Taking
program~(\ref{tag:ex1}) as an example again, the object invariant will
imply of the individual specifications of $T_1$ and $T_2$, that $r_1$
must equal $2$, and $r_2$ must equal $1$, if no threads interfered
with $T_1$ and $T_2$.




We can similarly employ abstract histories to specify
\code{flip2}. One way to do it is to notice that the value of the
shared counter $x$ will be changing as $0, 1, 0, 1, \ldots$, and
exactly \emph{two} of these values will be contributed by each call to
\code{flip2} made by some thread. We can depict a particular total
history $\hist_{\cal F}$ of the \code{flip2} structure as follows:
\[
%
\!\!\!\!\!\!\!\!
\begin{array}{r@{\ }c@{\ }c@{\ }c@{\ }c@{\ }c@{\ }c@{\ }c@{\ }c@{\ }l@{\ }}
  & & \boxed{T_1} & \graybox{\boxed{T_2}} & {\boxed{T_1}} & \graybox{\boxed{T_2}} & \graybox{\boxed{T_3}}
  & \graybox{\boxed{T_3}} 
  \\[5pt] 
  \hist_{\cal F} = \text{[} & ..., & 1, & \graybox{0}, & 
                                 1, & \graybox{0}, &  \graybox{1}, & \graybox{0}, &
                                                                          \ldots
                     &
                       \text{]}   
\\[-5pt]
&& \multicolumn{3}{l}{{\underbracket{\phantom{aaaaaaaaa}}_{T_1.\mathtt{flip2}}}}
\end{array}
\]
The two ``white'' contributions are made by thread $T_1$'s call to
\code{flip2}, while the rest (gray) are contributions by $T_1$'s
environment. Since the atomic \code{flip} operation returns the
\emph{complementary} (\ie, previous) value of the counter, the overall
result of $T_1$'s call in this case is $\bar{1} + \bar{1} = 0 + 0
=~0$.

The invariant for the \code{flip2} structure postulates the
interleaving 0/1-shape of the history and also ensures that the last
history entry is $x$'s current value. This will allow us to reason
about clients of \code{flip2}, such as~\eqref{tag:flip3}.
%
%
In the absence of interference, we can deduce that the two parallel
calls to \code{flip2} have contributed four \emph{consecutive} entries
to the history $\hist_{\cal F}$, with each thread contributing
precisely two of them. For each of the two calls, the result equals
the sum of the two complementary values for what the corresponding
thread has contributed to the history, hence, the overall sum
$r_1 + r_2$ is $2$.


\subsection{Hoare-style specifications for \texttt{exchange} and
  \texttt{flip2}}
\label{sec:hoare}

The above examples illustrate that subjectivity and object invariants
are two sides of the same coin. In tandem, they allow us to specify
threads individually, but also reason about thread combinations. We
emphasize that in our approach, the invariants are
\emph{object-specific} and \emph{provided by the user}. For example,
we can associate the invariant about twin entries with the exchanger
structure, but our method will not mandate the same invariant for
other structures for which it is not relevant. This is in contrast to
using a fixed correctness condition, such as linearizability, QC, or
CAL, which cannot be parametrized by user-defined
properties.\footnote{For example, linearizability does not allow users
  to declare history invariants on a per-object basis. The exchanger
  example motivated the introduction of the correctness condition
  CAL~\cite{Hemed-Rinetzky:PODC14}, which relaxes linearizability, and
  makes it somewhat more general in this respect, but still falls
  short of admitting user-defined invariants. \texttt{flip2} can be
  specified using a variation of QC~\cite{Jagadeesan-Riely:ICALP14},
  but we show that a similar property can be expressed via
  subjectivity and a user-defined invariant.}



Subjective histories can be encoded in our approach as \emph{auxiliary
  state}~\cite{Sergey-al:ESOP15,Owicki-Gries:CACM76}. Our Hoare
triples will specify how programs modify their histories, while the
invariants are declared as properties of a chunk of shared state
(\eg,~resource invariants of~\cite{Owicki-Gries:CACM76}). With the two
components, we will be able to describe the effects and results of
programs \emph{declaratively}, \ie, without exposing program
implementations.

\begin{figure*}
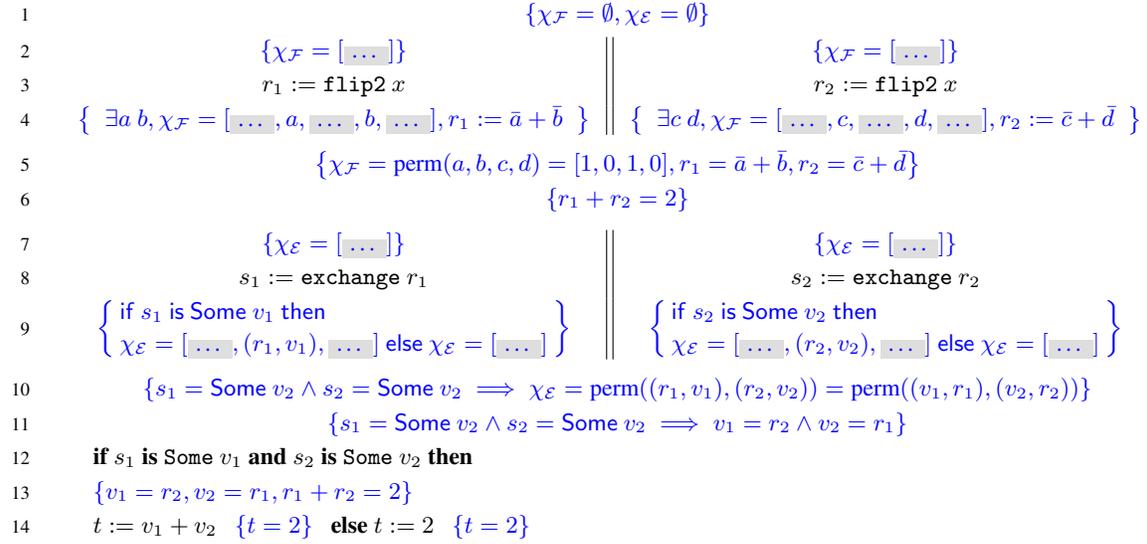

\centering
{\small{
\[
\begin{tabular}{r@{\ \ \ \ \ \ \ \ }c || c}
\Num{1} & \multicolumn{2}{c}{\specK{\{\hist_{\cal F} = \emptyset,  \hist_{\cal E} = \emptyset\}}}
\\[3pt]
\Num{2} & \specK{\{\hist_{\cal F} = [\graybox{\ldots}] \}}
&
\specK{\{\hist_{\cal F} = [\graybox{\ldots}] \}}
\\[2pt]
\Num{3} & $r_1 := \esc{flip2}~x$ & $r_2 := \esc{flip2}~x$
\\[2pt]
\Num{4} & \spec{
  \begin{array}{c}
    \exists a~b,
   \hist_{\cal F} = [\graybox{\ldots}, a, \graybox{\ldots}, b,
    \graybox{\ldots}], 
    r_1 := \bar{a} + \bar{b}
  \end{array}
}
&
\spec{
  \begin{array}{c}
    \exists c~d,
   \hist_{\cal F} = [\graybox{\ldots}, c, \graybox{\ldots}, d,
    \graybox{\ldots}], 
    r_2 := \bar{c} + \bar{d}
  \end{array}
}
\\ & \multicolumn{2}{c}{}\\[-5pt]
\Num{5} & \multicolumn{2}{c}{\spec{\hist_{\cal F} = \text{perm}(a, b, c, d) = [1, 0,
  1, 0], r_1 = \bar{a} + \bar{b}, r_2 = \bar{c} + \bar{d}}}
\\[2pt]
\Num{6} & \multicolumn{2}{c}{\spec{r_1 + r_2 = 2}}
\\ & \multicolumn{2}{c}{}\\[-5pt]
\Num{7} &
\specK{\{\hist_{\cal E} = [\graybox{\ldots}] \}} & 
\specK{\{\hist_{\cal E} = [\graybox{\ldots}] \}}
\\[2pt]
\Num{8} & $s_1 := \esc{exchange}~r_1$ & $s_2 := \esc{exchange}~r_2$
\\[2pt]
\Num{9} & 
  \spec{\!\!
  \begin{array}{l}
    \mathsf{if}\ s_1\ \mathsf{is}\ \mathsf{Some}\ v_1\ \mathsf{then}~\\[1pt]
    \hist_{\cal E} = [\graybox{\ldots}, (r_1, v_1), \graybox{\ldots}]
    ~\mathsf{else}\ \hist_{\cal E} = [\graybox{\ldots}]
  \end{array}
  \!\!}
&
  \spec{\!\!
  \begin{array}{l}
    \mathsf{if}\ s_2\ \mathsf{is}\ \mathsf{Some}\ v_2\ \mathsf{then}~\\[1pt]
    \hist_{\cal E} = [\graybox{\ldots}, (r_2, v_2), \graybox{\ldots}]
    ~\mathsf{else}\ \hist_{\cal E} = [\graybox{\ldots}]
  \end{array}
  \!\!}%
\\ & \multicolumn{2}{c}{}\\[-5pt]
\Num{10} &
\multicolumn{2}{c}{
\spec{s_1 = \mathsf{Some}\ v_2 \wedge s_2 = \mathsf{Some}\ v_2
  \implies
\hist_{\cal E} = \text{perm}((r_1, v_1), (r_2, v_2)) = \text{perm}((v_1, r_1),
  (v_2, r_2))
}}
\\[2pt]
\Num{11} &
\multicolumn{2}{c}{
\spec{s_1 = \mathsf{Some}\ v_2 \wedge s_2 = \mathsf{Some}\ v_2
  \implies
v_1 = r_2 \wedge v_2 = r_1
}}
\\[2pt]
\Num{12} &
\multicolumn{2}{l}
{$\kw{if}~s_1~\kw{is}~\esc{Some}~v_1~\kw{and}~s_2~\kw{is}~\esc{Some}~v_2~\kw{then}$}
\\[2pt]
\Num{13} &
\multicolumn{2}{l}
{\spec{v_1 = r_2,  v_2 = r_1, r_1 + r_2 = 2}}
\\[2pt]
\Num{14} &
\multicolumn{2}{l}
{$t := v_1 + v_2$~~~{\spec{t = 2}}~~~$ \kw{else}~t := 2$~~~{\spec{t = 2}}}
\end{tabular} 
\]
}}  
\caption{Verification of a concurrent client program using
  \code{exchange} and \code{flip2} in the absence of external
  interference.}
\label{fig:verif1}
\end{figure*}

A semi-formal and partial spec of \code{exchange} looks as follows,
with the white/gray parts denoting \emph{self}/\emph{other}
contributions to history, from the point of view of the thread being
specified (we postpone the full presentation until
Section~\ref{sec:exchanger}):
%
%
{\small{
\[
\tag{\arabic{tags}}\refstepcounter{tags}\label{tag:exsimpl} 
{\small{
\begin{array}{c}
\specK{\{\hist_{\cal E} = [\graybox{\ldots}] \}}\\[2pt]
\esc{exchange}\ v\\[2pt]
  \spec{\!\!
  \begin{array}{c}
    \mathsf{if}\ \res\ \mathsf{is}\ \mathsf{Some}\ w\ \mathsf{then}~\\[1pt]
    \hist_{\cal E} = [\graybox{\ldots}, (v, w), \graybox{\ldots}]
    ~\mathsf{else}\ \hist_{\cal E} = [\graybox{\ldots}]
  \end{array}
  \!\!}
\end{array}
}}
\]
}}
\hspace{-5pt}
The ellipsis ($\ldots$) stands for an existentially-quantified chunk
of the history.
%
%
%
%
%
The spec~\eqref{tag:exsimpl} says that a successful exchange adds an
entry $(v, w)$ to the \emph{self}-history (hence, the entry is
white). In the case of failed exchange, no entry is added. In the
complete and formal specification in Section~\ref{sec:exchanger}, we
will have to add a timing aspect, and say that the new entry appears
\emph{after} all the history entries from the precondition. We will
also have to say that no entries are removed from the \emph{other}
history (\ie, the exchanger cannot erase the behavior of other
threads), but we elide those details here.


%
%

The spec of \code{flip2} is defined with respect to history
$\hist_{\cal F}$:
\[
\tag{\arabic{tags}}\refstepcounter{tags}\label{tag:flipsimpl} 
{\small{
\begin{array}{c}
\specK{\{\hist_{\cal F} = [\graybox{\ldots}] \}}\\[2pt]
\esc{flip2}\ x\\[2pt]
  \spec{\!\!
  \begin{array}{c}
   \exists a~b, \hist_{\cal F} = [\graybox{\ldots}, a, \graybox{\ldots}, b,
    \graybox{\ldots}], \res = \bar{a} + \bar{b}
  \end{array}
  \!\!}
\end{array}
}}
\]
It says that the return value $\res$ is equal to the sum of binary
complements $\bar{a} + \bar{b}$ for the thread's two separate
\emph{self}-contributions to the history. Due to the effects of the
interference, the history entries $a$ and $b$ may be separated in the
overall history by the contributions of the environment, as indicated
by \graybox{\ldots} between them.

\subsection{Using subjective specifications in the client code}
\label{sec:clients}

The immediate benefit of using Hoare logic is that one can easily
reason about programs whose components use different object
invariants, whereas there is not much one can say about programs whose
components require different correctness conditions.
For example, Figure~\ref{fig:verif1} shows a proof sketch for a toy
program that uses both \code{exchange} and \code{flip2}.  As each of
these methods requires its own auxiliary history variable
($\hist_{\cal E}$ for the exchanger, and $\hist_{\cal F}$ for
\code{flip2}), the combined program uses both, but the proof simply
ignores those histories that are not relevant for any specific method
(\ie, we can ``frame'' the specs~\eqref{tag:exsimpl}
and~\eqref{tag:flipsimpl} wrt.~the histories of the objects that they
do not depend upon).

%

The program first forks two instances of \code{flip2}, storing the
results in $r_1$ and $r_2$ (line~4). Next, two new threads are forked,
trying to exchange $r_1$ and $r_2$ (line~8). The conditional (line~12)
checks if the exchange was successful, and if so, assigns the sum of
exchanged values to $t$ (line~14); otherwise $t$ gets assigned 2. We
want to prove via the specs~\eqref{tag:exsimpl}
and~\eqref{tag:flipsimpl}, that in the absence of external
interference on the \code{flip2}'s pointer $x$ and the exchanger, the
outcome is always $t = 2$.

\paragraph{Explaining the verification}

In addition to the absence of external interference, we assume that
the initial value of $x$ is $0$, and the initial \emph{self}-histories
for both \code{flip2} and \code{exchange} are empty (line~1).
%
%
%
%
%
Once the \code{flip2} threads are forked, we employ
spec~\eqref{tag:flipsimpl} for each of them, simply ignoring (i.e.,
framing out) $\hist_{\cal E}$, as this history variable does not apply
to them \code{flip2}. Upon finishing, the postconditions of
\code{flip2} in line~4 capture the relationship between the
contributions to the history $\hist_{\cal F}$ and the results $r_1$
and $r_2$ of the two calls.

Both postconditions in line~4 talk about the very same history
$\hist_{\cal F}$, just using different colors to express that the
contributions of the two threads are \emph{disjoint}: $a$ and $b$
being white in the left thread, implies that $a$ and $b$ are history
entries added by the left thread. Thus, they \emph{must} be gray in
the right thread, as they cannot overlap with the entries contributed
by the right thread. The right thread cannot explicitly specify in its
postcondition that $a$ and $b$ are gray, since the right thread is
unaware of the specific contributions of the left thread.
%

Dually, $c$ and $d$ being white in the right thread in line~4, implies
that they must be gray on the left. Thus, overall, in line~5, we know
that $\hist_{\cal F}$ contains all four entries in some permutation,
and in the absence of intereference, it contains no other entries but
these four. From the object invariant on $\hist_{\cal F}$ it then
follows that the entries are some permutation of $[1,0,1,0]$, which
makes their sum total $r_1 + r_2 = 2$. 
%


Similarly, we ignore $\hist_{\cal F}$ while reasoning about calls to
\code{exchange} via spec~\eqref{tag:exsimpl} (lines~7 and 9). As
before, we know that the entry $(r_1, v_1)$, which is white in the
left postcondition in line~9, must be gray on the right, and dually
for $(r_2, v_2)$. In total, the history $\hist_{\cal E}$ must contain
both of the entries, but, by the invariant, it must also contain their
twins. In the absence of any other interference, it therefore must be
that $(r_1, v_1)$ is a twin for $(r_2, v_2)$, \ie, $r_1 = v_2$ and
$r_2 = v_1$, as line~11 expresses for the case of a succesful
exchange.
%
%
The rest of the proof is then trivial.

The sketch relied on several important aspects of program verification
in FCSL: \emph{(i)} the invariants constraining $\hist_{\cal F}$ and
$\hist_{\cal E}$ were preserved by the methods, \emph{(ii)} upon
joining the threads, we can rely on the disjointness of history
contributions of the two threads, in order to combine the thread-local
views into a specification of the parent thread, and, \emph{(iii)} we
could guarantee the absence of the external interference.

The aspect \emph{(i)} is a significant component of what it means to
specify and verify a concurrent object. As we will show in
Sections~\ref{sec:exchanger} and~\ref{sec:counting}, defining a
sufficiently strong object invariant, and then proving that it is
indeed an invariant, \ie, that it is preserved by the implementation
of the program, is a major part of the verification challenge.
We will explain FCSL rules for \emph{parallel composition} and
\emph{hiding} in Section~\ref{sec:background}, justifying the
reasoning principles~\emph{(ii)} and \emph{(iii)}.

\subsection{Specifying non-linearizable objects in three steps}
\label{sec:three-steps-reas}

As shown by Sections~\ref{sec:hist}--\ref{sec:clients}, our method for
specifying and verifying non-linearizable concurrent objects and their
clients boils down to the following three systematic steps.

\paragraph{Step 1 (\S\ref{sec:hist}):} 

\emph{Define object-specific auxiliary state and its invariants. The
  auxiliary state will typically include a specific notion of abstract
  histories, recording whatever behavior is perceived as essential by
  the implementor of the object}.
%
%
To account for the variety of object-specific correctness conditions,
we do not fix a specific shape for the histories. We do not restrict
them to always record pairs of numbers (as in the exchanger), or
record single numbers (as in \code{flip2}). The only requirement that
we impose on auxiliary state in general, and on histories in
particular, is that the chosen type of auxiliary state is an instance
of the PCM algebraic structure~\cite{Sergey-al:ESOP15}, thus providing
an abstract, and user-defined, notion of \emph{disjointness} between
\emph{self}/\emph{other} contributions.

\paragraph{Step 2 (\S\ref{sec:hoare}):} 

\emph{Formulate Hoare-style specifications, para-metrized by
  interference, and verify them}.
This step provides a suitable ``interface'' for the methods of the
concurrent object, which the clients use to reason, without
knowing the details of the object and method implementations.
Naturally, the interface can refer to the auxiliary state and
histories defined in the previous step.
When dealing with non-linearizable objects in FCSL, it is customary to
formulate the spec in a subjective way (\ie, using
\emph{self}/\emph{other}, dually white/gray division between history
entries) so that the specification has a way to refer to the effects
of the interfering calls to the same object. 
The amount of interference can be later instantiated with more
specific information, once we know more about the context of
concurrent threads in which the specified program is being run.

\paragraph{Step 3 (\S\ref{sec:clients}):} 

\emph{Restrict the interference when using object specs for
  verification of clients}.
Eventually, thread-local knowledge about effects of individual clients
of one and the same object, should be combined into a cumulative
knowledge about the effect of the composition.
To measure this effect, one usually considers the object in a
\emph{quiescent} (interference-free) moment~\cite{Rinard:RACES}.
%
%
To model quiescent situations, FCSL provides a program-level
constructor for \emph{hiding}. In particular, $\esc{hide}\ e$ executes
$e$, but statically prevents other threads from interfering with $e$,
by making $e$'s auxiliary history invisible. Program $e$'s
\emph{other} contribution is fixed to be empty, thus modeling
quiescence.

\section{Verifying the Exchanger Implementation}
\label{sec:exchanger}

We now proceed with more rigorous development of the invariants and
specification for the exchanger data structure, necessary to verify
its real-world implementation~\cite{ExchangerClass}, which was so far
elided from the overview of the approach.


\newcommand{\Unmatched}{{\mathsf{U}}}
\newcommand{\Matched}[1]{{\mathsf{M}\ #1}}
\newcommand{\Retired}{{\mathsf{R}}}

The exchanger implementation is presented in ML-style pseudo-code in
Figure~\ref{fig:exchanger}. It takes a value $v\,{:}\,A$ and creates
an \emph{offer} from it (line 2). An offer is a pointer $p$ to two
consecutive locations in the heap.\footnote{In our mechanization, we
  simplify a bit by making $p$ point to a pair instead.}
%
The first location stores $v$, and the second is a ``hole'' which the
interfering thread tries to fill with a matching value. The hole is
drawn from the type
$\esc{hole}\,{=}\,\Unmatched\,{\mid}\,\Retired\,{\mid}\,\Matched
w$. Constructor $\Unmatched$ signals that the offer is unmatched;
$\Retired$ that the exchanger retired (\ie, withdrew) the offer, and
does not expect any matches on it; and $\Matched w$ that the offer has
been matched with a value $w$.

The global pointer $g$ stores the latest offer proposed for
matching. The exchanger proposes $p$ for matching by making $g$ point
to $p$ via the atomic compare-and-set instruction \code{CAS} (line
3). We assume that \code{CAS} returns the value read, which can be
used to determine if it failed or succeeded. If \code{CAS} succeeds,
exchanger waits a bit, then checks if the offer has been matched by
some $w$ (lines 6, 7). If so, $\esc{Some}\ w$ is returned (line
7). Otherwise, the offer is retired by storing $\Retired$ into its
hole (line 6). Retired offers remain allocated (thus, exchanger has a
memory leak) in order to avoid the ABA problem, as usual in many
concurrent structures~\cite{Herlihy-Shavit:08,Treiber:TR}.
If the exchanger fails to link $p$ into $g$ in line 3, it deallocates
the offer $p$ (line 10), and instead tries to match the offer $cur$
that is current in $g$. If no offer is current, perhaps because
another thread already matched the offer that made the \code{CAS} in
line 3 fail, the exchanger returns $\mathsf{None}$ (line
12). Otherwise, the exchanger tries to make a match, by changing the
hole of $cur$ into $\Matched v$ (line 14). If successful (line 16), it
reads the value $w$ stored in $cur$ that was initially proposed for
matching, and returns it. In any case, it unlinks $cur$ from $g$ (line
15) to make space for other offers.

{
\begin{figure}
\centering
\[
{\small{
\begin{array}{rl}
 \Num{1} & \esc{exchange}~(v : A) : \esc{option}~A~=~\{ 
\\ 
 \Num{2} & ~~~~ p \Asgn \esc{alloc}~(v, \Unmatched);\\
 \Num{3} & ~~~~ b \Asgn \esc{CAS}~(g, \esc{null}, p);\\
 \Num{4} & ~~~~ \kw{if}~~b~\esc{==}~\esc{null}~~\kw{then}\\
 \Num{5} & ~~~~ ~~~~ \esc{sleep}~(50);\\
 \Num{6} & ~~~~ ~~~~ x \Asgn \esc{CAS}~(p\esc{+}1, \Unmatched, \Retired);\\
 \Num{7} & ~~~~ ~~~~ \kw{if}~~x~~\kw{is}~~\Matched w~~\kw{then}~~\kw{return}~~(\esc{Some}~w)\\
 \Num{8} & ~~~~ ~~~~ \kw{else}~~\kw{return}~~\esc{None}\\
 \Num{9} & ~~~~ \kw{else}\\
\Num{10} & ~~~~ ~~~~ \esc{dealloc}~p;\\
\Num{11} & ~~~~ ~~~~ cur \Asgn \esc{read}~g;\\
\Num{12} & ~~~~ ~~~~ \kw{if}~~cur~\esc{==}~\esc{null}~~\kw{then}~~\kw{return}~{\esc{None}}\\
\Num{13} & ~~~~ ~~~~ \kw{else}\\
\Num{14} & ~~~~ ~~~~ ~~~~ x \Asgn \esc{CAS}~(cur\esc{+}1, \Unmatched, \Matched v);\\
\Num{15} & ~~~~ ~~~~ ~~~~ \esc{CAS}~(g, cur, \esc{null});\\
\Num{16} & ~~~~ ~~~~ ~~~~ \kw{if}~~x~~\kw{is}~~\Unmatched~~\kw{then}~~w\Asgn \esc{read}~cur;\kw{return}~(\esc{Some}\ w)\\
\Num{17} & ~~~~ ~~~~ ~~~~ \kw{else}~~\kw{return}~\esc{None}\}
\end{array}
}}
\]
\caption{Elimination-based exchanger procedure.}
\label{fig:exchanger}
\end{figure} 
}

\subsection{Step 1: defining auxiliary state and invariants}

%
%

To formally specify the exchanger, we decorate it with auxiliary
state. 
In addition to histories, necessary for specifying the observable
behavior, the auxiliary state is used for capturing the coherence
constraints of the actual implementation, \eg, with respect to memory
allocation and management of outstanding offers.
The state is subjective as described in Section~\ref{sec:overview}: it
keeps thread-local auxiliary variables that name the thread's private
state (\emph{self}), but also the private state of all other threads
combined (\emph{other}).
%

The subjective state of the exchanger for each thread in this example
consists of three groups of two components: (1) thread-private heap
$\heaps$ of the thread, and of the environment $\heapo$, (2) a set of
outstanding offers $\perms$ created by the thread, and by the
environment $\permo$, and (3) a time-stamped history of values
$\hists$ that the thread exchanged so far, and dually $\histo$ for the
environment. In Section~\ref{sec:overview}, we illustrated
subjectivity by means of histories, white we used white and gray
entries, respectively, to describe what here we name $\hists$ and
$\histo$, respectively. Now we see that the dichotomy extends beyond
histories, and this example requires the dichotomy applied to heaps,
and to sets of offers as well. In addition to \emph{self}/\emph{other}
components of heaps, permissions and histories, we also need shared
(aka.~\emph{joint}) state consisting of two components: a heap
$\heapj$ of storing the offers that have been made, and a map
$\pending$ of offers that have been matched, but not yet collected by
the thread that made them.

Heaps, sets and histories are all PCMs under the operation of disjoint
union, with empty heap/set/history as a unit. We overload the notation
and write $x\,{\mapsto}\,v$ for a singleton heap with a pointer $x$
storing value $v$, and $t\,{\mapsto}\,a$ for a singleton
history. Similarly, we apply disjoint union $\hunion$ and subset
$\subseteq$, to all three types uniformly.

We next describe how the exchanger manipulates the above variables.
First, $\heapj$ is a heap that serves as the ``staging'' area for the
offers. It includes the global pointer $g$. Whenever a thread wants to
make an offer, it allocates a pointer $p$ in $\heaps$, and then tries
to move $p$ from $\heaps$ into $\heapj$, simultaneously linking $g$ to
$p$, via the \code{CAS} in line~3 of Figure~\ref{fig:exchanger}.

Second, $\perms$ and $\permo$ are sets of offers (hence, sets of
pointers) that determine offer ownership. A thread that has the offer
$p \in \perms$ is the one that created it, and thus has the
\emph{sole} right to retire $p$, or to collect the value that $p$ was
matched with. Upon collection or retirement, $p$ is removed from
$\perms$.


Third, $\hists$ and $\histo$ are exchanger-specific histories, each
mapping a time-stamp (isomorphic to nats), to a pair of exchanged
values. A singleton history $t \mapsto (v, w)$ symbolizes that a
thread having this singleton as a subcomponent of $\hists$, has
exchanged $v$ for $w$ at time $t$.
%
%
As we describe below, the most important invariant of the exchanger is
that each such singleton is matched by a ``symmetric'' one to capture
that another thread has \emph{simultaneously} exchanged $w$ for
$v$. Classical linearizability cannot express this simultaneous
behavior, making the exchanger non-linearizable.

Fourth, $\pending$ is a map storing the offers that were matched, but
not yet acknowledged and collected. Thus,
$\mathsf{dom}\ \pending = \perms \hunion \permo$. A singleton entry in
$\pending$ has the form $p \mapsto (t, v, w)$ and denotes that offer
$p$, initially storing $v$, was matched at time $t$ with $w$. A
singleton entry is entered into $\pending$ when a thread on the one
end of matching, matches $v$ with $w$. Such a thread also places the
\emph{twin} entry $\twin{t} \mapsto (w, v)$, with inverted order of
$v$ and $w$, into its own private history $\hists$, where:
\[
\begin{array}{c}
\twin{t} = \left\{%
\begin{array}{ll}
t+1 & \mbox{if $t$ is odd}\\
t-1 & \mbox{if $t > 0$ and $t$ is even}
\end{array}\right.  
\end{array}
\]
For technical reasons, $0$ is not a valid time-stamp, and has no
distinct twin. The pending entry for $p$ resides in $\pending$ until
the thread that created the offer $p$ decides to ``collect'' it. It
removes $p$ from $\pending$, and simultaneously adds the entry $t
\mapsto (v, w)$ into its own $\hists$, thereby logically completing
the exchange. Since twin time-stamps are consecutive integers, a
history cannot contain entries \emph{between} twins.

Thus, two twin entries in the combined history including $\hists$,
$\histo$ and $\pending$, jointly represent a single exchange, as if it
occurred \emph{atomically}. \emph{Concurrency-aware}
histories~\cite{Hemed-Rinetzky:PODC14} capture this by making the ends
of an exchange occur as simultaneous events. We capture it via twin
time-stamps. More formally, consider
$\hist = \hists \hunion \histo \hunion \mygather{\pending}$. Then, the
exchanger's main invariant is that $\hist$ always contains matching
twin entries:
%
\[
\tag{\arabic{tags}}\refstepcounter{tags}\label{tag:exchanging} 
t \mapsto (v, w) \subseteq \hist \iff \twin t \mapsto (w, v) \subseteq \hist
\]
Here $\mygather{\pending}$ is the collection of all the entries in
$\pending$. That is, $\mygather{\emptyset} = \emptyset$, and
$\mygather {p \mapsto (t, v, w) \hunion \pending'} = t \mapsto (v, w)
\hunion \mygather{\pending'}$.

In our implementation, we prove that atomic actions, such as
\code{CAS}, preserve the invariant, therefore, the whole program,
being just a composition of actions, doesn't violate~it.


\subsection{Step 2: Hoare-style specification of Exchanger}

We can now give the desired formal Hoare-style spec.
\[
\tag{\arabic{tags}}\refstepcounter{tags}\label{tag:exchangespec} 
{\small{
\begin{array}{c}
\specK{\{\heaps = \emptyset, \perms = \emptyset, \hists = \emptyset, \gist \subseteq \histo \hunion \mygather{\pending}\}}\\
\esc{exchange}\ v\\
  \spec{\!\!
  \begin{array}{c}
\heaps = \emptyset, \perms = \emptyset, \gist \subseteq
  \histo \hunion \mygather{\pending}, \hbox{}\\[1pt]
\mathsf{if}\ \res\ \mathsf{is}\ \mathsf{Some}\ w\ \mathsf{then}~\\[1pt]
\exists t\ldot \hists = t \mapsto (v, w), \mathsf{last} (\gist) < t, \twin{t}
~\mathsf{else}\ \hists = \emptyset 
  \end{array}
  \!\!}
\end{array}
}}
\]
The precondition says that the exchanger starts with the empty private
heap $\heaps$, set of offers $\perms$ and history $\hists$; hence by
framing, it can start with any value for these
components.\footnote{Framing in FCSL is similar to that of separation
  logic, allowing extensions to the initial state that remain
  invariant by program execution. In FCSL, however, framing applies to
  any PCM-valued state component (\eg, heaps, histories, \etc.),
  whereas in separation logic, it applies just to heaps.} The logical
variable $\gist$ names the initial history of all threads,
$\histo \hunion \mygather{\pending}$, which may grow during the call,
thus, we use subset instead of equality to make the precondition
stable under other threads adding
 entries to $\histo$ or $\pending$.

In the postcondition, the self heap $\heaps$ and the set of offers
$\perms$ didn't change. Hence, if $\mathtt{exchange}$ made an offer
during its execution, it also collected or retired it by the end.
The history $\gist$ is still a subset of the ending value for $\histo
\hunion \mygather{\pending}$, signifying that the environment history
only grows by interference. We will make a crucial use of this part of
the spec when verifying a client of the exchanger in
Section~\ref{sec:cal}.

If the exchange fails (\ie, $\mathsf{res}$ is $\mathsf{None}$), then
$\hists$ remains empty.  If it succeeds (either in line 7 or line 16
in Figure~\ref{fig:exchanger}), \ie, if the result $\mathsf{res}$ is
$\mathsf{Some}\ w$, then there exists a time-stamp $t$, such that
self-history $\hists$ contains the entry $t \mapsto (v, w)$,
symbolizing that $v$ and $w$ were exchanged at time $t$.

Importantly, the postcondition implies, by
invariant~(\ref{tag:exchanging}), that in the success case, the twin
entry $\twin t \mapsto (w, v)$ must belong to $\histo \hunion
\mygather{\pending}$, \ie, \emph{another} thread matched the exchange
(this was made explicit by the spec~\eqref{tag:exsimpl}).
Moreover, the exchange occurred \emph{after} the call to
$\mathsf{exchange}$: whichever $\gist$ we chose in the pre-state, both
$t$ and $\twin t$ are larger than the last time-stamp in $\gist$.

The proof outline for the exchanger is available in
Appendix~\ref{app:exch}.
In Section~\ref{sec:cal}, after introducing necessary FCSL background,
we will illustrate \textbf{\emph{Step 3}} of our method and show how
to employ the subjective Hoare spec~\eqref{tag:exchangespec} for
modular verification of a concurrent client.


\section{Background on FCSL}
\label{sec:background}

In order to formally present \textbf{\emph{Step 3}} of our method, we
first need to introduce some important parts of FCSL.

A Hoare specification in FCSL has the form $\spec{P}\ e\ \spec{Q} @
\rcon$. $P$ and $Q$ are pre- and postcondition for partial
correctness, and $\rcon$ defines the \emph{shared resource} on which
$e$ operates. 
%
%
The latter is a state transition system describing the invariants of
the state (real and auxiliary) and atomic operations that can be
invoked by the threads that simultaneously operate on that state.
%
%
We elide the transition system aspect of resources here, and refer
to~\cite{Nanevski-al:ESOP14} for detailed treatment.

An important secondary role of a resource is to declare the variables
that $P$ and $Q$ may scope over. For example, in the case of
exchanger, we use the variables $\heaps, \perms, \hists$, $\heapo,
\permo, \histo$, and $\heapj, \pending$.
The mechanism by which the variables are declared is as
follows. Underneath, a resource comes with only three variables:
$a_\lcl$, $a_\env$ and $a_\joint$ standing for abstract self state,
other state, and shared (joint) state, but the user can pick their
types depending on the application. In the case of exchanger, $a_\lcl$
and $a_\env$ are triples containing a heap, an offer-set and a
history. The variables we used in Section~\ref{sec:overview} are
projections out of such triples: $a_\lcl\,{=}\,(\heaps, \perms,
\hists)$, and $a_\env\,{=}\,(\heapo, \permo, \histo)$. Similarly,
$a_\joint\,{=}\,(\heapj, \pending)$.

It is essential that $a_\lcl$ and $a_\env$ have a common type
exhibiting the algebraic structure of a PCM, under a partial binary
operation $\hunion$.
%
PCMs give a way, generic in $\rcon$, to define the
inference rule for parallel composition.
\[
\tag{\normalsize \arabic{tags}}\refstepcounter{tags}\label{eq:parrule}
{\small{
\begin{array}{c}
\specK{\{P_1\}}\ e_1\ \specK{\{Q_1\}} @ \rcon \quad \specK{\{P_2\}}\ e_2\ \specK{\{Q_2\}} @ \rcon\\[2pt]
\hline\\[-7pt]
\specK{\{P_1 \circledast P_2\}}\ e_1 \parallel e_2\ \specK{\{[\res.1/\res]Q_1 \circledast [\res.2/\res]Q_2\}} @ \rcon
\end{array}
}}
\]
Here, $\circledast$ is defined as follows.
\[
\begin{array}{c}
(P_1 \circledast P_2)(a_\lcl, a_\joint, a_\env) \iff \exists x_1~x_2\ldot a_\lcl = x_1 \hunion x_2, \hbox{}\\
 P_1 (x_1, a_\joint, x_2 \hunion a_\env), P_2 (x_2, a_\joint, x_1 \hunion a_\env)
\end{array}
\]
Thereby, when a parent thread forks $e_1$ and $e_2$, then $e_1$
becomes part of the environment for $e_2$, and vice-versa. This is so
because the \emph{self} component $a_\lcl$ of the parent is split into
$x_1$ and $x_2$; $x_1$ becomes the \emph{self} part of $e_1$, but
$x_2$ is added to the \emph{other} part $a_\env$ of $e_1$ (and
symmetrically for $e_2$).

To reason about quiescent moments, we use one more constructor of
FCSL: \emph{hiding}. The program $\mathsf{hide}\ e$ operationally
executes $e$, but logically installs a resource within the scope of
$e$. In the case of the exchanger, $\mathsf{hide}\ e$ starts only with
private heaps $\heaps$ and $\heapo$, then takes a chunk of heap out of
$\heaps$ and ``installs'' an exchanger in this heap, allowing the
threads in $e$ to exchange values. $\mathsf{hide}\ e$ is
\emph{quiescent} wrt.~exchanger, as the typechecker will prevent
composing $\mathsf{hide}\ e$ with threads that want to exchange values
with $e$.

The auxiliaries $\perms, \hists$, $\permo, \histo$, and $\heapj,
\pending$, belonging to the exchanger (denoted as resource $\cal E$)
are visible within $\mathsf{hide}$, but outside, only $\heaps$
persists (denoted as a resource $\cal P$ for private state).  We elide
the general hiding rule~\cite{Nanevski-al:ESOP14}, and just show the
special case for the exchanger.

\[
\tag{\normalsize \arabic{tags}}\refstepcounter{tags}\label{eq:ehide}
{\small{
\begin{array}{c}
\specK{\{P\}}\ e\ \specK{\{Q\}} @ \cal E\\[2pt]
\hline\\[-7pt]
\specK{\{\heaps = \Phi_1(\heapj), \Phi_1(P)\}}\ \mathsf{hide}~e\ \specK{\{\exists \Phi_2\ldot \heaps = \Phi_2(\heapj), \Phi_2(Q)\}} @ \cal P
\end{array}
}}
\]

Read bottom-up, the rule says that we can install the exchanger $\cal
E$ in the scope of a thread that works with $\cal P$, but then we need
substitutions $\Phi_1$ and $\Phi_2$, to map variables of $\cal E$
($\heaps, \perms, \hists$, \etc) to values expressed with variables
from $\cal P$ ($\heaps$ and $\heapo$). $\Phi_1$ is an initial such
substitution (user provided), and the rule guarantees the existence of
an ending substitution $\Phi_2$. The substitutions have to satisfy a
number of side conditions, which we elide here for brevity. The most
important one is that \emph{other} variable $a_\env = (\heapo, \permo,
\histo)$ is fixed to be the PCM unit (\ie,~a triple of empty
sets). Fixing $a_\env$ to unit captures that $\mathsf{hide}$ protects
$e$ from interference.

At the beginning of $\mathsf{hide}~e$, the private heap equals the
value that $\Phi_1$ gives to $\heapj$ ($\heaps = \Phi_1(\heapj)$). In
other words, the $\mathsf{hide}$ rule takes the private heap of a
thread, and makes it shared, \ie, gives it to the $\heapj$ component
of $\cal E$. Upon finishing, $\mathsf{hide}~e$ makes $\heapj$ private
again.
%


%
%

In the subsequent text we elide the resources from specs.

\section{Verifying Exchanger's Client}
\label{sec:cal}
\newcommand{\ts}{\mathit{ts}}
\newcommand{\vvs}{\mathit{vs}}
\newcommand{\acc}{\mathit{ac}}
\newcommand{\ws}{\mathit{ws}}
\newcommand{\sorted}[1]{\mathsf{sorted}\ #1}


We next illustrate how the formally specified exchanger from
Section~\ref{sec:exchanger} can be used by real-world client programs,
and how the \emph{other} component, asserted by the spec to satisfy
$\gist \subseteq \histo \hunion \mygather{\pending}$, is crucial for
their verification.
We emphasize that the proof of the client does not see the
implementation details, which are hidden by the
spec~\eqref{tag:exchangespec}.

While simple, our client is realistic, and has been used
in~\code{java.util.concurrent}~\cite{ExchangerClass}. It is defined as
follows. First, the exchanger loops until it exchanges the value.
\vspace{-2pt}
\[
\vspace{-2pt}
{\small{
\begin{array}{rl}
& \esc{exchange'}~(v : A) : A = \{\\[1pt]
&  ~~~~ w' \Asgn \esc{exchange}~v;\\[1pt]
&  ~~~~
  \kw{if}~~w'~~\kw{is}~~\esc{Some}~w~~\kw{then}~~\kw{return}~w~~\kw{else}~~\esc{exchange'}~v~\}
\end{array}
}}
\]
Next, $\esc{exchange'}$ is iterated to exchange a sequence in order,
appending the received matches to an accumulator.
%
\[
{\small{
\begin{array}{rl}
& \esc{ex\_seq}~(\vvs, \acc : \esc{seq}~A) : \esc{seq}~A = \{\\[1pt]
& ~~~~ \kw{if}~~\vvs~~\kw{is}~~v{::}\vvs'~~\kw{then}\\[1pt]
& ~~~~ ~~~~ w \Asgn \esc{exchange'}~v;~~\esc{ex\_seq}~(\vvs', \esc{snoc}~\acc~w)\\[1pt]
& ~~~~ \kw{else}~~\kw{return}~\acc~\}
\end{array}
}}
\]
Our goal is to prove, via~\eqref{tag:exchangespec},
that the parallel composition
\[
e = \esc{ex\_seq}~(\vvs_1, \esc{nil}) \parallel \esc{ex\_seq}~(\vvs_2, \esc{nil})
\]
exchanges $\vvs_1$ and $\vvs_2$, \ie,~returns the pair $(\vvs_2,
\vvs_1)$. This holds only under the assumption that $e$ runs without
interference (\ie, quiescently), so that the two threads in $e$ have
no choice but to exchange the values between themselves. 

We make the quiescence assumption explicit using the FCSL $\hide$
constructor, as described in Section~\ref{sec:background}.
%
%
Thus, we establish the following Hoare triple:
\[
\tag{\arabic{tags}}\refstepcounter{tags}\label{tag:hidespec} 
{\small{
\!\!\!\!\!
\begin{array}{c}
\specK{\{\heaps = g \mapsto\mathsf{null}\}}~~\hide~~e~~\specK{\{g \in
  \mathsf{dom}~\heaps, \res = (\vvs_2, \vvs_1)\}} 
\end{array}
}}
\]
It says that we start with a heap where $g$ stores $\mathsf{null}$,
and end with a possibly larger heap (due to the memory leak), but with
the result $(\vvs_2, \vvs_1)$. The auxiliaries $\perms, \permo$,
$\gists, \gisto$, $\heapj, \pending$ are visible inside $\hide$, but
outside, only $\heaps$ persists.



\paragraph{Explaining the verification.}
We illustrate the verification by listing the specs of selected
subprograms. First, the spec of $\esc{exchange'}$ easily derives
from~(\ref{tag:exchangespec}) by removing the now-impossible failing
case.
%
\[
{\small{
\begin{array}{c}
\specK{\{\heaps = \emptyset, \perms = \emptyset, \hists = \emptyset, \gist \subseteq \histo \hunion \mygather{\pending}\}}\\[2pt]
\esc{exchange'}\ v\\[2pt]
\spec{\!\!
\begin{array}{c}
\heaps = \emptyset, \perms = \emptyset, \gist \subseteq \histo \hunion \mygather{\pending}, \\[1pt]    
\exists t\ldot \hists = t \mapsto (v, \res), \mathsf{last} (\gist) < t, \twin{t}
\end{array}
\!\!}
\end{array}
}}
\]
Next, $\esc{ex\_seq}$ has the following spec:
\[
{\small{
\begin{array}{c}
\specK{\{\heaps = \emptyset, \perms = \emptyset, \hists = \emptyset\}}\\[2pt]
\mathtt{ex\_seq}~(vs, \mathsf{nil})\\[2pt]
\spec{\!\!\!
\begin{array}{c}
\exists \ts\ldot \heaps = \emptyset, \perms = \emptyset, 
\hists = \mathsf{zip}~\ts~\vvs~\res,
\\[1pt]
\mathsf{grows\_notwins}~\ts, 
\mathsf{zip}~\overline{\ts}~\res~\vvs \subseteq \histo  \hunion \mygather{\pending}  
\end{array}
\!\!\!}
\end{array}
}}
\]
Here, $\ts$ is a list of time-stamps, and
$\mathsf{zip}\,\ts\,\vvs\,\ws$ joins up the singleton histories
$t\,{\mapsto}\,(v, w)$, for each $t$, $v$, $w$ drawn, in order, from
the lists $\ts$, $\vvs$, $\ws$.
%
%
The spec says that at the time-stamps from $\ts$, $\esc{ex\_seq}$
exchanged the elements of $\vvs$ for those of $\esc{res}$. That $\ts$
is increasing and contains no twins, follows from the spec of
$\esc{exchange'}$ which says that the time-stamps $t$ and $\bar t$
that populate $\ts$ and $\overline{\ts}$, are larger than anything in
$\gist$, and thus only grow with iteration.
%
%
From the same postcondition, it follows that $\histo \hunion
\mygather{\pending}$ contains all the twin exchanges, by
invariant~(\ref{tag:exchanging}), as commented in
Section~\ref{sec:overview} about the spec for $\esc{exchange}$.

Next, by the FCSL parallel composition rule
(Section~\ref{sec:background}):
\[
{\small{
\begin{array}{c}
\specK{\{\heaps = \emptyset, \perms = \emptyset, \hists = \emptyset\}}\\[2pt]
\mathsf{ex\_seq}~(\vvs_1, \mathsf{nil}) 
\parallel
\mathsf{ex\_seq}~(\vvs_2, \mathsf{nil}) \\[1pt]
\spec{\!\!\!
\begin{array}{c}
\exists \ts_1~\ts_2\ldot \mathsf{grows\_notwins}~{\ts_1}, \mathsf{grows\_notwins}~{\ts_2},\\
 \heaps = \emptyset, \perms = \emptyset, \hists = \mathsf{zip}~\ts_1~\vvs_1~\res.1 \hunion \mathsf{zip}~\ts_2~\vvs_2~\res.2,\\
 \mathsf{zip}~\overline{\ts_1}~\res.1~\vvs_1 \subseteq \mathsf{zip}~\ts_2~\vvs_2~\res.2 \hunion \histo \hunion\mygather{\pending}, \\
 \mathsf{zip}~\overline{\ts_2}~\res.2~\vvs_2 \subseteq \mathsf{zip}~\ts_1~\vvs_1~\res.1 \hunion \histo \hunion\mygather{\pending}. 
\end{array}
\!\!\!}
\end{array}
}}
\]
%
%
To explain: $ts$ and $\esc{res}$ from the left and right
$\esc{ex\_seq}$ threads become $ts_1$, $ts_2$, $\esc{res}.1$ and
$\esc{res}.2$, respectively. The values of each \emph{self} component
$\heaps$, $\perms$, $\hists$ from the two threads are joined into the
\emph{self} component of the composition. At the same time, the
\emph{other} component $\histo$ of the left (resp.~right) thread
equals the sum of $\hists$ of the right (resp.~left) thread, and the
$\histo$ of the composition.  This formalizes the intuition that upon
forking, the left thread becomes part of the environment for the right
thread, and vice-versa.

The postcondition says that the self history of $e$ contains both
$\mathsf{zip}\,\ts_1\,\vvs_1\,\res.1$ and
$\mathsf{zip}\,\ts_2\,\vvs_2\,\res.2$. Thus, $\vvs_1$ is exchanged for
$\res.1$, and $\vvs_2$ for $\res.2$. But we further want to derive
$\res.1\,{=}\,\vvs_2$ and $\res.2\,{=}\,\vvs_1$, \ie, the lists are
exchanged \emph{for each other}, in the absence of interference.

We next explain how this desired property follows for $\hide~e$, from
the two inequalities in $e$'s postcondition
\begin{align}
\mathsf{zip}~\overline{\ts_1}~\res.1~\vvs_1\,&\subseteq&\!\!\!\!{\mathsf{zip}}~\ts_2~\vvs_2~\res.2\,&\,{\hunion}\,\histo\,{\hunion}\,\mygather{\pending}, \tag{\arabic{tags}}\refstepcounter{tags}\label{tag:x}\\
\mathsf{zip}~\overline{\ts_2}~\res.2~\vvs_2\,&\subseteq&\!\!\!\!{\mathsf{zip}}~\ts_1~\vvs_1~\res.1\,&\,{\hunion}\,\histo\,{\hunion}\,\mygather{\pending}. \tag{\arabic{tags}}\refstepcounter{tags}\label{tag:y}
\end{align}
Notice that $(\ref{tag:x})$ and $(\ref{tag:y})$ are ultimately
instances of the conjunct $\gist \subseteq \histo \hunion
\mygather{\pending}$ that was part of the
specification~(\ref{tag:exchangespec}), thereby justifying the use of
subjective \emph{other} variables.

We know that $\mathsf{dom}\ \pending\,{=}\,\perms \hunion \permo$
(from Section~\ref{sec:overview}), that $\perms\,{=}\,\emptyset$ (from
$e$'s postcondition), and that by hiding,
$\permo\,{=}\,\histo\,{=}\,\emptyset$. Thus, towards deriving the
postcondition of $\hide~e$, we simplify $(\ref{tag:x})$ and
$(\ref{tag:y})$ into:
\begin{align*}
\mathsf{zip}~\overline{\ts_1}~\res.1~\vvs_1 \subseteq \mathsf{zip}~\ts_2~\vvs_2~\res.2\\
\mathsf{zip}~\overline{\ts_2}~\res.2~\vvs_2 \subseteq \mathsf{zip}~\ts_1~\vvs_1~\res.1
\end{align*}
Because $\ts_1$ and $\ts_2$ are increasing lists of time-stamps, and
contain no twins, the above implies $\ts_2 = \overline{\ts_1}$. Hence:
%
\[
\mathsf{zip}~\overline{\ts_1}~\res.1~\vvs_1 = \mathsf{zip}~\ts_2~\vvs_2~\res.2
\]
and thus $\res.1\,{=}\,\vvs_2$, $\vvs_1\,{=}\,\res.2$. We omit the
remaining technical argument that explains how the heap $\heapj$, with
the pointer $g$, is folded into $\heaps$, which ultimately
obtains~\eqref{tag:hidespec}.

\section{Specifying Counting Networks}
\label{sec:counting}

We now show how to use subjective histories to specify another class
of non-linearizable objects---\emph{counting networks}.
Counting networks are a special case of \emph{balancing networks}
introduced by Aspnes \etal~\cite{Aspnes-al:JACM94}, themselves
building on sorting networks~\cite{Ajtai-al:STOC83}, aimed to
implement concurrent counters in a way free from synchronization
bottlenecks.
The key idea is to decompose the workload between \emph{several}
counters, so that each of them is responsible for a disjoint set of
values. A thread trying to increment first approaches the
\emph{balancer}, which is a logical ``switch'' that ``directs'' the
thread, \ie, provides it with the address of the counter to increment.
The balancers make counting networks' operations
\emph{non-linearizable}, as in the presence of interference the
results of increments might be observed out of order.
%
{
\begin{figure}
\begin{tabular}{c@{\ \ \ \ \ \ }c}
\begin{minipage}[c]{2.5cm}
\includegraphics[width=2.1cm]{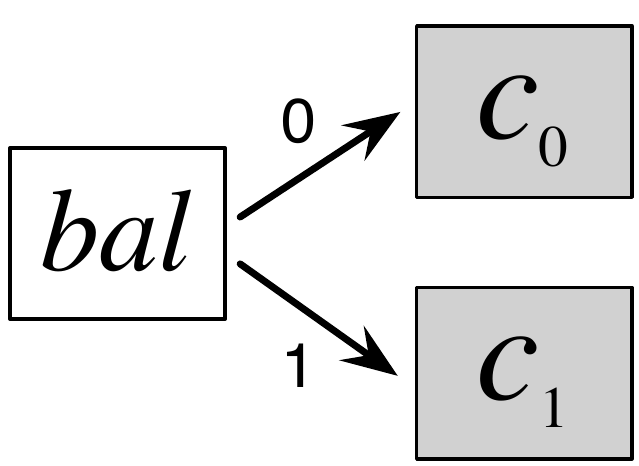} 
\end{minipage}
&
\begin{minipage}[l]{4.9cm}
\centering
{\small{
\[
\begin{array}{rl}
\Num{1} & \esc{getAndInc()} : \esc{nat}~=~\esc{\{}  \\[2pt] 
\Num{2} & ~~~~ b \Asgn \esc{flip(}\bal\esc{)};\\[2pt]
\Num{3} & ~~~~ \res \Asgn \esc{fetchAndAdd2(}c_b\esc{)};\\[2pt]
\Num{4} & ~~~~ \kw{return}~\res~\esc{\}}
\end{array}
\]
}}
\end{minipage} 
\end{tabular}
\vspace{-10pt}  
\caption{Simple counting network}
\label{fig:counter-fig} 
\end{figure}
}

Figure~\ref{fig:counter-fig} presents a schematic outline and a
pseudo-code implementation of a counting network with a single
balancer.
The implementation contains three pointers: the balancer $\bal$, which
stores either 0 or 1, thus directing threads to the shared pointers
$c_0$ or $c_1$, which count the even and odd values,
respectively. Threads increment by calling \code{getAndInc}, which
works as follows. It first atomically changes the bit value of the
balancer via a call to atomic operation \code{flip} (line 2). The
\code{flip} operation returns the \emph{previous} value $b$ of the
balancer as a result, thus determining which of the counters, $c_0$ or
$c_1$, should be incremented. The thread proceeds to atomically add 2
to the value of $c_b$ via \code{fetchAndAdd2} (line 3). The old value
of $c_b$ is returned as the result of the procedure.\footnote{In the
  counting network from Figure~\ref{fig:counter-fig}, the balancer
  itself might seem like a contention point. However, the \code{flip}
  operation is much less expensive than \code{CAS} as a
  synchronization mechanism. The performance can be further improved
  by constructing a \emph{diffracting tree} of several
  balancers~\cite[\S 12.6]{Herlihy-Shavit:08}, but we do not consider
  diffracting trees here.}

Assuming that $c_0$ and $c_1$ are initialized with $0$ and $1$, it is
easy to see that in a single-threaded program, the network will behave
as a conventional counter; that is, consecutive invocations of
\code{getAndInc} return consecutive nats.
However, in the concurrent setting, \code{getAndInc} may return
results out of order, as follows. 
%

\vspace{3pt}
\begin{example}
\label{ex:t1t2}
Consider two threads, $T_1$ and $T_2$ operating on the network
initialized with $\bal\,{\mapsto}\,0$, $c_b\,{\mapsto}\,b$. $T_1$
calls \code{getAndInc} and executes its line~2 to set $\bal$ to 1. It
gets suspended, so $T_2$ proceeds to execute lines~2 and~3, therefore
setting $\bal$ back to $0$ and returning $1$. While $T_1$ is still
suspended, $T_2$ calls \code{getAndInc} again, gets directed to $c_0$,
and returns 0, after it has just returned 1.
\end{example}
\vspace{3pt}

\noindent

This out-of-order behavior, however, is not random, and can be
precisely characterized as a function of the number of threads
operating on the
network~\cite{Afek-al:OPODIS10,Jagadeesan-Riely:ICALP14}. In the rest
of this section and in Section~\ref{sec:qclients}, we show how to
capture such bounds in the spec using auxiliary state of (subjective)
histories in a client-sensitive manner. As a form of road map, we list
the desired requirements for the spec of \code{getAndInc},
adapting the design goals of the criteria, such as QC, QQC and
QL~\cite{Aspnes-al:JACM94,Afek-al:OPODIS10,Jagadeesan-Riely:ICALP14},
which we will proceed to verify formally, following \textbf{\emph{Step
    1}} and \textbf{\emph{Step 2}} of our approach, and then employ in
client-side reasoning via \textbf{\emph{Step 3}}:
\vspace{2pt}
\begin{itemize}

\item \textbf{R1:} Two different calls to \code{getAndInc}
  should return distinct results (\emph{strong concurrent
    counter semantics}).

\item \textbf{R2:} The results of calls to \code{getAndInc},
  separated by a period of quiescence (\ie, absence of interference),
  should appear in their sequential order (\emph{quiescent
    consistency}).

\item \textbf{R3:} The results of two sequential calls $C_1$ and
  $C_2$, in a single thread should be out of order by no more than $2\
  N$, where $N$ is the number of interfering calls that overlap with
  $C_1$ and $C_2$ (\emph{quantitative quiescent consistency}).

\end{itemize}


\subsection{Step 1: counting network's histories and invariants}
\label{sec:counting-intuition}

To formalize the necessary invariants, we elaborate the counting
network with auxiliary state: \emph{tokens} (isomorphic to nats) and
novel \emph{interference-capturing histories}.

A \emph{token} provides a thread that owns it with the right to
increment an appropriate counter~\cite{Aspnes-al:JACM94}. In our
example, a thread that performs the \code{flip} in line 2 of
\code{getAndInc} will be awarded a token which it can then spend to
execute \code{fetchAndAdd2}.
Thus, any individual token represents a ``pending'' call to
\code{getAndInc}, and the set of unspent tokens serves as a bound on
the out-of-order behavior that the network exhibits. We introduce
auxiliary variables for the held tokens: $\tkns$ keeps the tokens
owned by the \emph{self} thread, with its \emph{even} and \emph{odd}
projections $\tkns^0$ and $\tkns^1$, such that $\tkns = \tkns^0
\hunion \tkns^1$, administering access to $c_0$ and $c_1$,
respectively. Similarly, $\tkno$, featuring the same projections,
keeps the tokens owned by the \emph{other} thread.  We abbreviate
$\tkn^i = \tkns^i \hunion \tkno^i$ for $i=0,1$.  
%
%

Figure~\ref{fig:chist} illustrates a network with three \emph{even}
tokens: $x^0, y^0, z^0 \in \tkn^0$, held by threads that will
increment $c_0$, and one \emph{odd} token $u^1 \in \tkn^1$, whose
owner will increment $c_1$.
%

A \emph{history} of the counting network is an auxiliary finite map,
consisting of entries of the form $t \mapsto (\tknh, z)$.  Such an
entry records that the value $t$ has been written into an appropriate
counter ($c_0$ or $c_1$, depending on the parity of $t$), at the
moment when $\tkn^0$ and $\tkn^1$ held values of $\tknh$'s even/odd
projections $\tknh^0$ and $\tknh^1$, respectively. Moreover, in order
to write $t$ into a counter, the token $z$ was spent by the thread. We
will refer to $z$ as the \emph{spent} token. Notice that the entries
in the history contain tokens held by both \emph{self} and
\emph{other} threads. Thus, a history captures the behavior of a
thread subjectively, \ie, as a function of the interfering threads'
behavior.

Similarly to tokens, network histories are represented by the
auxiliary variables $\hists$, tracking counter updates (even and odd)
performed by the \emph{self} thread, and dually $\histo$ for the
\emph{other} thread. We abbreviate $\hist^i = \hists^i \hunion
\histo^i$ for $i = 0,1$.

Figure~\ref{fig:chist} illustrates a moment in network's history and
how it relates to the state of the counters. Only $0$ has been written
to $c_0$ so far (upon initialization), hence $\hist^0$ only contains
an entry for $t = 0$ (we ignore at the moment the \emph{contents} of
the history entries). On the other hand, $\hist^1$ has entries for $1$
and $3$, because after initialization, one thread has increased $c_1$.
The gray boxes indicate that $0$ and $3$ are the current values of
$c_0$ and $c_1$, and thus also the latest entries in $\hist^0$ and
$\hist^1$, respectively. In particular, these values will be returned
by the next invocations of \code{fetchAndAdd2}. The dashed boxes
correspond to the entries to be contributed by the currently running
threads holding tokens $x^0$, $y^0$, $z^0$, $u^1$.
%

{
\setlength{\belowcaptionskip}{-15pt} 
\begin{figure}
\centering
\includegraphics[width=8.2cm]{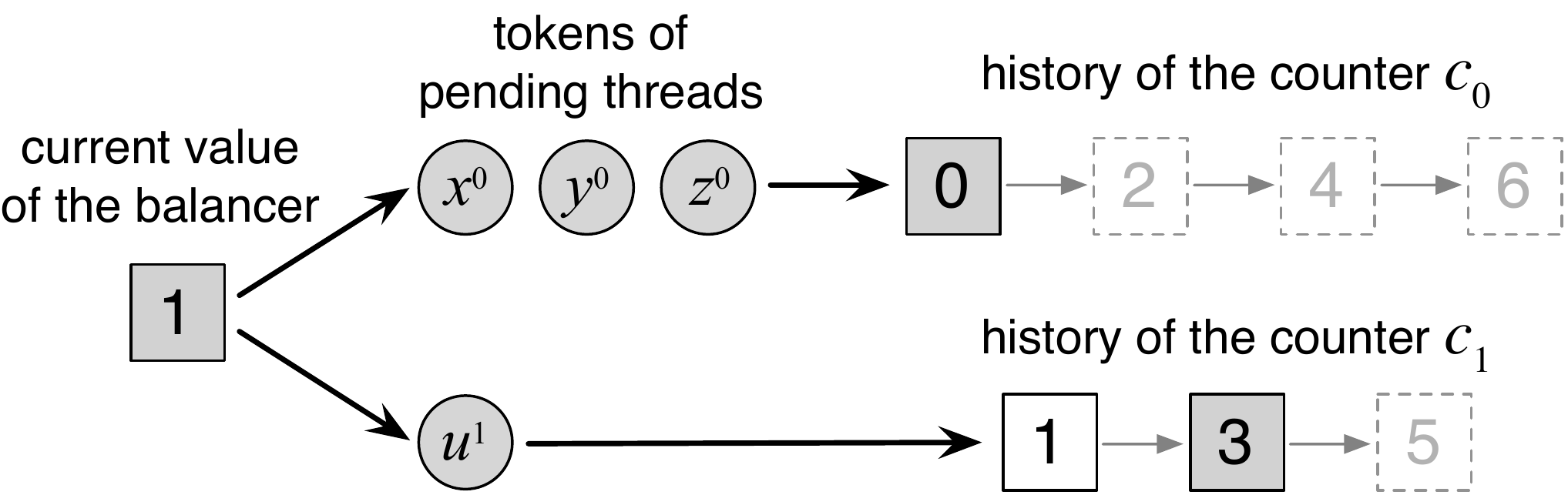}      
\caption{Tokens and histories of the simple network}
\label{fig:chist}
\end{figure}
}


In addition to $\tkn$ and $\hist$ which come in flavors private to
\emph{self} and \emph{other} threads, we require the following shared
variables: (1) $\heapj$ for the joint heap of the network, and (2)
$b_\joint$, $n^0_\joint$ and $n^1_\joint$ for the contents of $\bal$,
$c_o$ and $c_1$, respectively.

\paragraph{Invariants of the counting network}
\label{sec:count-netw-invar}

The main invariant of the network relates the number of tokens, the
size of histories and the value of the balancer:

\[
\tag{\normalsize{\arabic{tags}}}\refstepcounter{tags}\label{cn:si} 
|\hist^0| + |\tkn^0| =
|\hist^1| + |\tkn^1| + b_\joint
\]

The equation formalizes the intuition that out-of-order anomalies of
the counting network appear if one of the two counters is too far
ahead of the other one.
The invariant~(\ref{cn:si}) provides a bound on such a situation. One
counter can get ahead temporarily, but then there must be a number of
threads waiting to spend their tokens on the other counter. Thus, the
other counter will eventually catch up.

The approaches such as quiescent and quantitative quiescent
consistency describe this situation by referring to the number of
\emph{unmatched} call events in an event
history~\cite{Derrick-al:FM14,Jagadeesan-Riely:ICALP14}. In contrast,
we formalize this property via auxiliary state: the sets of tokens
$\tknh$ recorded in the entry for the number $t$ determine the
environment's capability to add new history entries, and thus ``run
ahead'' or ``catch up'' after $t$ has been returned.
%
%
The other invariants of the counting network are as follows:
\begin{enumerate}[label=(\roman*)]

%


\item\label{cn:state} $\heapj = \bal \mapsto b_\joint \hunion c_0 \mapsto n^0_\joint
  \hunion c_1 \mapsto n^1_\joint$.

\item\label{cn:hvalid} The histories contain disjoint time-stamps. 
 

\item\label{cn:ci} 
  The history $\hist^0$ (resp. $\hist^1$) contains \emph{all} even
  (resp. odd) values in $[0, n^0_\joint]$ (resp. $[1, n^1_\joint]$).
%
%
    This ensures that $n^0_\joint$ and $n^1_\joint$ are the last
    time-stamps in $\hist^0$ and $\hist^1$, respectively.

\item\label{cn:ti}  
  $\tkn^0$, $\tkn^1$ and $\Tomb~(\hists \hunion \histo)$ contain
  mutually disjoint tokens, where $\Tomb~(t \mapsto (\tknh, z) \hunion
  \hist') = \{z\} \hunion \Tomb~\hist'$, and $\Tomb~\emptyset =
  \emptyset$. In other words, a spent token never appears among the
  ``alive'' ones (\ie, in $\tkn^0 \hunion \tkn^1$).


\item\label{cn:ti1}
  $t \mapsto (\tknh, z) \subseteq \hists \hunion \histo
  \implies z \in \tknh$. \\[-7pt]

\item\label{cn:ai} 
For any $t$, $\tknh$, $z$: \\[-7pt]
{\small
  \begin{itemize}
  \item   $t \hpts (\tknh, z) \subseteq \hist^0 \implies t + 2\ |\tknh
    \cap \tkn^0| < n^1_\joint + 2\ |\tknh \cap \tkn^1| + 2$, and \\[-7pt]
  \item
    $t \hpts (\tknh, z) \subseteq \hist^1 \implies t + 2\ |\tknh \cap 
    \tkn^1| < n^0_\joint + 2\ |\tknh \cap \tkn^0|
    + 2$.
  \end{itemize}
}
\end{enumerate}
\vspace{5pt}
 
\noindent
The invariant~\ref{cn:ai} provides quantitative information about the
network history by relating the actual ($n^0_\joint$, $n^1_\joint$)
and the past ($t$) counter values, via the current amount of
interference ($\tkn$) and the snapshot interference ($\tknh$).
To explain~\ref{cn:ai}, we resort to the intuition provided by the
following equality, which, however, being \emph{not quite valid},
cannot be used as an invariant, as we shall see. Focusing on the
first clause in~\ref{cn:ai}, if
$t \mapsto (\tknh, z) \subseteq \hist^0$, then,
intuitively:
%
%
%
{\small{
\[
t + 2\ |\tknh^0 \setminus \tkn^0 | + 2\ |\tknh \cap \tkn^0| =
n^1_\joint + 2\ |\tknh \cap \tkn^1| + (2 b_\joint - 1)
\]}}
%
%
\noindent
The equality says the following. When $t$ is snapshot from $c_0$ and
placed into the history $\hist^0$, the set of outstanding even tokens
was $\tknh^0$. By the present time, $c_0$ has been increased
$|\tknh^0 \setminus \tkn^0|$ times, each time by $2$, thus
$n^0_\joint = t + 2\ |\tknh^0 \setminus \tkn^0|$. What is left to add
to $c_0$ to reach the \emph{period of quiescence}, when no threads
interfere with us, is $2\ |\tknh \cap \tkn^0|$. Similar reasoning
applies to $c_1$. It is easy to see at the period of quiescence, $c_0$
and $c_1$ differ by $2 b_\joint - 1$; that is, the counter pointed to
by $\bal$ is behind by $1$. However, the equality is invalid, as
$b_\joint$ can be read off only in the present, whereas the
``intuitive'' reasoning behind the equality requires a value of
$b_\joint$ from a quiescent period \emph{in the future}. Hence, in
order to get a valid property, we bound $2 b_\joint - 1$ by 2. For
simplicity, we even further weaken the bounds by dropping
$|\tknh^0 \setminus \tkn^0|$ to obtain~\ref{cn:ai}; as it will turn
out, even such a simpler bound will suffice for proving
\textbf{R1}--\textbf{R3}.



%
%
%

\paragraph{Allowed changes in the counting network}
\label{sec:count-netw-prot}

The state of the counting network (auxiliary and real) can be changed
in two possible ways by concurrent threads. These changes formalize
the way the atomic operations \code{flip} and \code{fetchAndAdd2} from
Figure~\ref{fig:counter-fig}~(b) work with auxiliary state.
\emph{Flipping} alters the bit value $b_\joint$ of $\bal$ to the
complementary one, $1 - b_\joint$.
It also generates a token $z$ (of parity $b_\joint$) and stores it
into $\tkns$. The token is fresh, \ie, distinct from all alive and
spent tokens in $\tkns \hunion \tkno \hunion{\Tomb~(\hists
  \hunion \histo)}$.
\emph{Incrementation} spends a token $z$ from $\tkns$, and depending
on its $i$, it atomically increases the value $n^i_\joint$ of $c_i$ by
two, while simultaneously removing $z$ from $\tkns$ (thus, the
precondition is that $z \in \tkns$). It also adds the entry
$(n^i_\joint + 2) \hpts (\tkn^0 \hunion \tkn^1, z^i)$ to $\hists$,
thus snapshoting the values of $\tkn^0$ and $\tkn^1$.
It is easy to check that both these allowed changes preserve the
state-space invariants~(\ref{cn:si}), \ref{cn:state}--\ref{cn:ai}, and
that their effect on real state (with auxiliary state erased) are
those of \code{flip} and \code{fetchAndAdd2}.

\subsection{Step 2: a Hoare spec for \texttt{getAndInc}}
\label{sec:spec-gaa}

Figure~\ref{fig:qspec} provides a Hoare-style spec for
\code{getAndInc}, verified in our proof scripts. We use the logical
variable $\ikn$ and its variants to range over token sets, and $\gist$
to range over histories.

\begin{figure}
\[
\tag{\normalsize \arabic{tags}}\refstepcounter{tags}\label{eq:qc-spec}
{\small
\!\!\!\!\!\!\!\! 
\begin{array}{c}
  \spec{\!\!
  \begin{array}{c}
    \tkns = \emptyset,
    \hists = \gists,
    \gisto \subseteq \histo,\\[2pt]
    \ikno \subseteq \tkno \hunion (\Tomb~\histo \setminus
    \Tomb~\gisto),
    \Ic{\gisto}{\ikno}
  \end{array}
  \!\!}
  \\\\[-6pt]
  \texttt{getAndInc()}
  \\[3pt]
  \spec{\!\!\!
  \begin{array}{c}
    \exists \iknh~z \ldot \tkns = \emptyset, 
    \hists = \gists \hunion (\res + 2) \hpts (\iknh, z), 
    \\[2pt]
    \gisto \subseteq \histo, \ikno \subseteq \tkno \hunion (\Tomb~\histo \setminus \Tomb~\gisto), 
    \\[2pt]
    \last~(\gists \hunion \gisto) < 
    \res + 2 + 2~|\iknh \cap \ikno|, 
    \\[2pt]
    \happrox~(\gists \hunion \gisto)~\res~\iknh~z,
     \Ic{\gisto}{\ikno}
  \end{array} 
  \!\!\!} 
\end{array}
}
\]
\caption{Hoare-style spec of a simple counting network.}
\label{fig:qspec}
\end{figure}

The precondition starts with an empty token set ($\tkns = \emptyset$),
and hence by framing, any set of tokens. The initial self-history
$\hists$ is set to an arbitrary $\gists$.\footnote{Alternatively, we
  could have also taken $\hists = \emptyset$, but the clients will
  require generalizing to $\hists = \gists$ by the FCSL's frame
  rule~\cite{Sergey-al:ESOP15}. To save space and simplify the
  discussion, we immediately frame \wrt the auxiliary $\hists$. Our
  examples do not require such client-side framing \wrt~$\tkns$.} The
precondition records the \emph{other} components of the initial state
as follows. First, $\gisto$ names (a subset of) $\histo$, to make it
stable under interference, as in Section~\ref{sec:overview}. Next, we
use $\ikno$ to name the (subset of) initially live tokens
$\tkno$. However, as $\tkno$ may shrink due to other threads spending
tokens, simply writing $\ikno \subseteq \tkno$ is unstable. Instead,
we write $\ikno \subseteq \tkno \hunion (\Tomb~\histo \setminus
\Tomb~\gisto)$ to account for the tokens spent by other threads as
well. The set $\tkno \hunion (\Tomb~\histo \setminus \Tomb~\gisto)$
only grows under interference, as new live tokens are generated, or
old live tokens are spent, making the inclusion of $\ikno$ stable.
Indeed, one cannot take \emph{any} arbitrary $\gisto$ and $\ikno$ to
name the \emph{other} components of the initial state. Therefore, we
constrain these two variables by the invariant $\ic$, that relates
them to the \emph{self-}components of the actual state and to each
other according to the
invariants~\ref{cn:hvalid}--\ref{cn:ai}.\footnote{That is, $\gisto$
  and $\ikno$ take the role of $\histo$ and $\tkno$ in
  invariants~\ref{cn:hvalid}--\ref{cn:ai}, with
  $n^i_\joint = \last~(\hists \hunion \gisto)^i$. The formal
  definition of $\ic$ is in our proof scripts.} This is natural,
since, as we will see in Section~\ref{sec:qclients}, all clients
instantiate $\gisto$ and $\ikno$ with the \emph{other}-components of
the actual pre-state, respecting~\ref{cn:hvalid}--\ref{cn:ai}.

%
%
%
%
%

The postcondition asserts that the final token set $\tkns$ is also
empty (\ie, the token that \code{getAndInc} generates by \code{flip},
is spent by the end). The history $\hists$ is increased by an entry
$(\res + 2) \hpts (\iknh, z)$, corresponding to writing the value of
the result (plus two) into one of the network's counters, snapshoting
the tokens of that moment into $\iknh$, and spending the token $z$ on
the write. $\gisto$ is a subset of the new value of $\histo$, and
$\ikno$ is a subset of the new value of $\tkno \hunion (\Tomb~\histo
\setminus \Tomb~\gisto)$, by the already discussed stability.

The next inequality describes where the entry for $\res + 2$ is placed
\wrt~the pre-state history $\gist = \gists \hunion \gisto$. $\gist$
may have gaps arising due to out-of-order behavior of the network, and
$\res + 2$ may fill one such gap. However, there is a bound on how far
$\res$ (and hence $\res+2$) may be from the tail of $\gist$. We
express it as a function of $\ikno$ and $\iknh$, derived from the
bounds in~\ref{cn:ai}, taking $\res + 2$ for $t$ and
over-approximating the instant value $n_{\joint}^i$ of the
incremented counter via $\last~(\gists \hunion \gisto)$. The
inequality weakens the invariant~\ref{cn:ai}, making it hold for even
and odd entries by moving $2~|\iknh \cap \ikno^i|$ (for $i = 0,1$) to
the right side of $<$ and joining them, since $\ikno^0 \cap \ikno^1 =
\emptyset$.

Finally, the predicate $\happrox$ provides more bounds that we will
need in the proofs of the client code's properties.
\[ 
\tag{\normalsize \arabic{tags}}\refstepcounter{tags}\label{eq:happrox}
\!\!\!\!\!
{\small{
\begin{array}{l}
\!\!\!\!
\happrox~\gist~\res~\iknh~z \eqdef \hbox{}
\iknh \subseteq \tkno \hunion (\Tomb~\histo) \hunion
  \set{z},\\[2pt]
~~ \forall t~\ikn \ldot t \hpts (\ikn, -) \subseteq \gist \Rightarrow
  z \notin \ikn,~  t < \res + 2 + 2 \ (|\iknh \cap \ikn|)
\end{array}
}}
\]
When instantiated with $\gist = \gists \hunion \gisto$, $\happrox$
says the following. The token set $\iknh$ snapshot when $\res+2$ was
committed to history, is a subset of all the tokens in post-state,
including the live ones ($\tkno$), and spent ones ($\Tomb~\histo
\hunion \{z\}$).
Moreover, if $t$ is an entry in $\gist$, with contents $(\ikn, -)$,
then: (1) $z \notin \ikn$, because $z$ is a token generated when
\code{getAndInc} executed \code{flip}. Hence, $z$ is fresh \wrt~any
token-set from the pre-state history $\gist$; and (2) $t$ and $\ikn$
satisfy the same bounds \wrt~$\res+2$, as those described for the last
history entry and~$\ikno$.


%


\paragraph{How will the spec~\eqref{eq:qc-spec} be used?}

The clause $\hists\,{=}\,\gists \hunion (\res+2)\,{\mapsto}\,-$ of
\eqref{eq:qc-spec}, in conjunction with the invariant~\ref{cn:hvalid},
ensures that any two calls to \code{getAndInc}, sequential or
concurrent, yield different history entries, and hence different
results. This establishes~\textbf{R1}, which we will not discuss
further.

The inequality on $\last~(\gists \hunion \gisto)$ will provide
for~\textbf{R2} in client reasoning. To see how, consider the
particular case when $\ikno$ is empty, \ie, the pre-state is
quiescent. In that case, the intersection with $\iknh$ is empty, and
we can infer that $\res + 2$, is larger than either counter's value in
the pre-state. As we shall see in Section~\ref{sec:qclients}, this
captures the essence of QC.

Finally, the predicate $\happrox$~\eqref{eq:happrox} establishes a
bound for the ``out-of-order'' discrepancy between the result $\res$
and any value $t$ committed to the history \emph{in the past}, via
$2~|\iknh \cap \ikn|$. We will further bound this value using the size
of $\iknh$, and the inclusion $\iknh \subseteq \tkno \hunion
\Tomb~\histo$ from~\eqref{eq:happrox}. These bounds will ultimately
enable us to derive the requirement~\textbf{R3}.

\section{Verifying Counting Network's Clients}
\label{sec:qclients}


Following \textbf{\emph{Step 3}} of our verification method, we now
illustrate requirements \textbf{R2} and \textbf{R3} from the previous
section via two different clients which execute two sequential calls
to \code{getAndInc}. Both clients are higher-order, \ie, they are
parametrized by subprograms, which can be ``plugged in''.
The first client will exhibit a quiescence between the two calls, and
we will prove that the call results appear in order, as required by
\textbf{R2}. The second client will experience interference of a
program with a $N$ concurrent calls to \code{getAndInc}, and we will
derive a bound on the results in terms of $N$, as required by
\textbf{R3}.

Both our examples will rely on the general mechanism of hiding,
presented in Section~\ref{sec:background}, as a way to logically restrict the
interference on a concurrent object, in this case, a counting network,
in a lexically-scoped way.
To ``initialize'' the counting network data structure, we provide the
starting values for the shared heap ($h_0$) and for the history
($\gist_0$), assuming that the initial set of tokens is empty:
%
%
\[
\tag{\normalsize \arabic{tags}}\refstepcounter{tags}\label{eq:hide2}
{\small{
\begin{array}{r@{\ }c@{\ }l}
\heap_0 & \eqdef & \bal \hpts 0 \hunion c_0 \hpts 0 \hunion c_1 \hpts 1     
\\[2pt]
\gist_0 & \eqdef & \set{0 \hpts (\set{0}, 0), 1 \hpts (\set{1}, 1)}
\end{array}
}}
\]
That is, $\gist_0$ provides the ``default'' history for the initial
values 0 and 1 of $c_0$ and $c_1$, with the corresponding tokens
represented by numbers 0 and 1.  As always with hiding, the
postcondition of the hidden program will imply that $\tkno$ and
$\histo$ are both empty, as there is no interference at the end.


\subsection{Exercising quiescent consistency}
\label{sec:qc-client}

\begin{figure}
\centering
\[
{\small{
\!\!\!\!\!\!\!\!
\begin{array}{c}
  \spec{\!\!
  \begin{array}{c}
    \tkns = \emptyset,
    \hists = \gists,
    \gisto \subseteq \histo, \Ic{\gisto}{\ikno}, \\[2pt]
    \ikno \subseteq \tkno \hunion (\Tomb~\histo \setminus \Tomb~\gisto)
  \end{array}
  \!\!}
\\\\[-5pt]
  \begin{tabular}{c || c}
   $\esc{getAndInc()}$ & ${\small{e_i}}$ 
\end{tabular}
\\\\[-5pt]
~~~~\spec{\!\!
\begin{array}{c}
  \exists \iknh~\gist_i \ldot  
  \tkns = \emptyset\aand \hists = \gbm{\gists \hunion \gist_i \hunion (\res.1 + 2) \hpts (\iknh, -)},\\[1pt]
  \gisto \subseteq \histo\aand \ikno \subseteq \tkno \hunion
  (\Tomb~\histo \setminus \Tomb~\gisto), \Ic{\gisto}{\ikno},\\[1pt]
  \last~(\gists \hunion \gisto)  < \gbm{\res.1} + 2 +
  2~|\iknh \cap \ikno|
\end{array}
\!\!} 
\end{array}
}}  
\]
\caption{Parallel composition of \code{getAndInc} and~$e_i$ in~\eqref{eq:eqc}.}
  \label{fig:example1} 
\end{figure}

Our first client is the following program~$\eqc$:
\[
\tag{\normalsize \arabic{tags}}\refstepcounter{tags}\label{eq:eqc}
{\small{
\begin{array}{ll} 
\Num{1} & (\res_1, -) \Asgn (\esc{getAndInc()} ~||~ e_1) \esc{;} \\[1pt]
\Num{2} & (\res_2, -) \Asgn (\esc{getAndInc()} ~||~ e_2) \esc{;} \\[1pt]
\Num{3} &  \kw{return}~(\res_1, \res_2) 
\end{array}
}}
\]
Each of the calls to \esc{getAndInc} interferes with either $e_1$ or
$e_2$, but in the absence of \emph{external} interference, the
quiescent state is reached between the lines 1 and 2. Hence, after
executing $\hide~\eqc$, it should be $\res_1 < \res_2$, following
\textbf{R2}.

The programs $e_1$ and $e_2$ can invoke \code{getAndInc} and modify
the counters concurrently with the two calls of $\eqc$, which we
capture by giving both the following generic spec:
\[
\tag{\normalsize \arabic{tags}}\refstepcounter{tags}\label{eq:eispec}
{\small
\!\!\!\!\!\!\!\! 
\begin{array}{c}
  \spec{~
  \hists = \emptyset\aand
  \tkns = \emptyset\aand
   \ikn \subseteq \tkno \hunion \Tomb~\histo
  ~}
  \\[1pt]
  e_i
  \\[1pt]
  \spec{\!\!\!
  \begin{array}{c}
    \exists \gist_i \ldot \hists = \gist_i\aand 
    \tkns = \emptyset\aand 
    \ikn \subseteq \tkno \hunion \Tomb~\histo
  \end{array} 
  \!\!\!} 
\end{array}
}
\]
The postcondition allows for a number of increments via calls to
\esc{getAndInc}, which is reflected in the addition $\gist_i$ to
$\hists$. However, all such calls are required to be \emph{finished}
by the end of $e_i$ ($\tkns = \emptyset$). As customary by now, we use
the logical variable $\ikn$ to name the initial set of \emph{other}
tokens.

Figure~\ref{fig:example1} provides a spec for each of the parallel
compositions in the program~\eqref{eq:eqc}, proved via the
corresponding FCSL inference rule for parallel
composition~\eqref{eq:parrule}.
The spec is very similar to~\eqref{eq:qc-spec} with the differences
highlighted via gray boxes: (a) the self-history $\hists$ is increased
by $e_i$'s contribution $\gist_i$ in addition to the entry, introduced
by \code{getAndInc}, (b) the result of the parallel composition is a
pair, but we only constrain its first component $\res.1$, resulting
from the left subprogram. We also drop the last conjunct with
$\happrox$ from~\eqref{eq:qc-spec}, which we won't require for this
example.

Next, we use the spec from Figure~\ref{fig:example1} to specify and
verify the program $\eqc$, so far \emph{assuming} external
interference.
\[
\!\!\!
{\small{
\begin{array}{c}
\!\!\!\!\!
\spec{\!\!
  \begin{array}{c}
    \mbox{Fig.~\ref{fig:example1}'s precondition with $\gists := \gist_0$, $\gisto :=
      \histo$, and $\ikno := \tkno$}
  \end{array}
  \!\!}
  ~\comm{P}
  \\\\[-6pt]
  (\res_1, -) \Asgn (\esc{getAndInc()} ~||~ e_1) \esc{;}
  \\[3pt]
\!\!\!\!{{
\spec{\!\!\!\!
\begin{array}{c}
 \exists \gist_1\ldot \tkns = \emptyset\aand \hists = \gists',~\ldots
\\[2pt]
\mbox{where 
 $\gbm{\gists' = \gist_0 \hunion
\gist_1 \hunion (\res_1 + 2)\mapsto -}$, $\gisto := \histo$ and $\ikno :=
\tkno$} 
  \end{array}
\!\!\!\!}
}}
\\\\[-5pt]
(\res_2, -) \Asgn (\esc{getAndInc()} ~||~ e_2) \esc{;}      
\\[3pt]
\spec{\!\!\!\!
\begin{array}{c}
\exists \gist_1~\gist_2~\iknh \ldot     
\tkns = \emptyset \aand 
\gbm{\ikno \subseteq \tkno \hunion (\Tomb~\histo \setminus \Tomb~\gisto)},\\[1pt]
\gbm{\last~(\gists' \hunion \gisto) < \res_2 + 2 + 2~|\iknh \cap \ikno|},~\ldots 
\end{array}
\!\!\!\!\!}~\comm{Q}
\\\\[-7pt]
~~~~~~~~~~~\kw{return}~(\res_1, \res_2); ~\comm{=: \res} 
\\[2pt]
\spec{~Q(\res.1/\res_1, \res.2/\res_2)~} 
\end{array}
}} 
\]
We start by instantiating the logical variables $\gists$, $\gisto$ and
$\ikno$ from Figure~\ref{fig:example1} with $\gist_0$, \emph{current}
$\histo$ and $\tkno$, respectively, naming the obtained precondition
$P$.
In the following assertion we focus on the clauses constraining
$\tkns$ and $\hists$. To verify the second call, we instantiate
$\gists$, $\gisto$ and $\ikno$ from Figure~\ref{fig:example1} with
$\gists' = \gist_0 \hunion \gist_1 \hunion (\res_1 + 2)\mapsto -$,
\emph{current} $\histo$ and $\tkno$, correspondingly, obtaining the
postcondition, which we name~$Q$.

The inequality in the postcondition $Q$ gives the boundary on the
out-of-order position of $\res_2$ with respect to the \emph{last}
value in the history captured in between the two parallel
compositions. The boundary is given via the size of intersection of
the two sets of tokens: snapshot ($\iknh$) and ``alive'' between the
calls ($\ikno$).
Now, to ensure the absence of external interference, we consider the
program $(\hide~\eqc)$.
By the general property of hiding (Section~\ref{sec:background}), we
know that at the final state there is no interference, hence $\tkno =
\emptyset$ and $\histo = \emptyset$ in $Q$.
Therefore, from the set inclusion on $\ikno$ in $Q$ (the grayed part),
we deduce that $\ikno = \emptyset$.
As a consequence, the intersection $\iknh \cap \ikno = \emptyset$, so
from the inequality we obtain
\[
\tag{\normalsize \arabic{tags}}\refstepcounter{tags}\label{eq:tada1}
%
\begin{array}{c}
 \last~(\gists' \hunion \gisto) < \res.2 + 2
\end{array}
\hfill
\]
%
%
But $\gists'$ is defined as $(\res.1+2)\mapsto -~\hunion\ldots$,
hence, $\res.1 + 2 \in \mathsf{dom}\ \gists'$, and thus $\res.1 + 2
\le \mathsf{last}\ \gists'$. Even more:
\[
\tag{\normalsize \arabic{tags}}\refstepcounter{tags}\label{eq:tada2}
%
\begin{array}{c}
\res.1 + 2 \le \last~(\gists' \hunion \gisto).
\end{array}
\hfill
\]
From~\eqref{eq:tada1} and~\eqref{eq:tada2} follows the result
\textbf{R2}: $\res.1 < \res.2$.

\subsection{Proving quantitative bounds}
\label{sec:qqc-client}

We next show how the spec~\eqref{eq:qc-spec} also obtains quantitative
bounds on the out-of-order anomalies in terms of a number of running
threads in the following program $\eqqc$:
\[
\tag{\normalsize \arabic{tags}}\refstepcounter{tags}\label{eq:eqqc}
{\small{
\begin{tabular}{l || l}
$
\begin{array}{ll} 
\Num{1} & \res_1 \Asgn \esc{getAndInc();} \\[1pt]
\Num{2} & \res_2 \Asgn \esc{getAndInc();}  \\[1pt]
\Num{3} & \kw{return}~(\res_1, \res_2)
\end{array}
$  
&
$~~~e$
\end{tabular} 
}}
\]
The $e$'s spec says that the \emph{number} of calls to \esc{getAndInc}
in~$e$ (\ie, the size of interference $e$ exhibits) is some fixed $N$:
\[
\tag{\normalsize \arabic{tags}}\refstepcounter{tags}\label{eq:espec}
{\small
\!\!\!\!\!\!\!\!\!  
\begin{array}{c}
  \spec{
  \tkns = \emptyset,
  \hists = \gists }
~  e
~  \spec{\!\!\!
  \begin{array}{c}
    \exists \gist \ldot 
    \tkns = \emptyset,
    \hists = \gists \hunion \gist,
    |\gist| = N
  \end{array} 
  \!\!\!} 
\end{array}
}
\]
Our goal is to prove that in the absence of external interference for
$\eqqc$, $\res_1 < \res_2 + 2 \ N$ (requirement \textbf{R3}).

\begin{figure}
\centering
\[
\!\!\!
{\small{
\begin{array}{c}
  \spec{~
    \mbox{{\normalsize{\eqref{eq:qc-spec}'}}s precondition with $\gists := \gist_0$, $\gisto :=
      \histo$, and $\ikno := \tkno$}~}
\\\\[-6pt]
\res_1 \Asgn \esc{getAndInc();}
\\[3pt]
\spec{\!\!
\begin{array}{c}
   \exists \ikn \ldot 
   \tkns = \emptyset, 
   \hists = \gists',
   \ldots
   \\[2pt]
   \mbox{where $\gbm{\gists' = \gist_0 \hunion (\res_1 + 2) \hpts
       (\ikn, -)}$} 
  \end{array}
  \!\!}%
\\\\[-5pt] 
\res_2 \Asgn \esc{getAndInc();}
\\[3pt]
\spec{\!\!\!
\begin{array}{c}
  \exists \iknh~z \ldot     
  \happrox (\gists' \hunion \gisto)~\res_2~\iknh~z, \ldots
\end{array}
\!\!\!}
\\\\[-5pt]
\spec{\!\!\!
\begin{array}{c}
  \exists \iknh~z \ldot     
  \gbm{\iknh \subseteq \tkno \hunion (\Tomb~\histo) \hunion \set{z}},
  z \notin \ikn,\\[2pt] 
  \res_1 + 2 < \res_2 + 2 + 2~|\iknh \cap \ikn|
\end{array}
\!\!\!}
\\\\[-5pt]  
~~~~~~~~~~~~~~~
\kw{return}~(\res_1, \res_2) ~\comm{=: \res}
\\\\[-5pt]
\spec{\!\!\!
\begin{array}{c}
  \res.1 < \res.2 + 2 \ |\tkno \hunion \Tomb~\histo| 
  %
  \end{array}
  \!\!\!} 
\end{array}
}} 
\]
\caption{Proof outline of sequential composition in~\eqref{eq:eqqc}.}
\label{fig:proof2}
\end{figure}

We first verify the sequential composition of the two calls
in~\eqref{eq:eqqc}; the proof outline is in
Figure~\ref{fig:proof2}. 
%
%
As previously, we start by instantiating the logical variables
$\gists$, $\gisto$ and $\ikno$ from spec~\eqref{eq:qc-spec} with
$\gists$, $\histo$ and $\tkno$, respectively. In the assertion,
resulting by of the first \esc{getAndInc}, we keep only the clauses
involving $\tkns$ and $\hists$, dropping the rest.
To verify the second \esc{getAndInc} call, we instantiate $\gists$,
$\gisto$ and $\ikno$ with $\gists' = \gists \hunion (\res_1+2) \mapsto
(\ikn, -)$, current $\histo$ and $\tkno$.

In the postcondition of the second call to \esc{getAndInc}, we focus
on the $\happrox~(\gists' \hunion \gisto)~\res_2~\iknh~z$ clause,
where $\iknh$ is the set of tokens snapshot when contributing
$\res_2+2$.
Unfolding the definition of $\happrox$ from~\eqref{eq:happrox}, we
obtain $\iknh \subseteq \tkno \hunion \Tomb~\histo
\hunion\{z\}$. Also, using $(\res_1 +2)\mapsto (\ikn, -)$ in the
implication that the unfolding obtains, we get $z \notin \ikn$ and
\[
\tag{\normalsize \arabic{tags}}\refstepcounter{tags}\label{eq:le0}
{\small{\res_1 + 2 < \res_2 + 2 + 2~|\iknh \cap \ikn|
}}
\]
Now we use the following trivial fact to simplify.
\vspace{8pt}
\begin{lemma}
\label{lm:intersect2}
If $z \in \iknh$ and $z \notin \ikn$, then $|\iknh \cap \ikn| \le
|\iknh| - 1$.
\end{lemma}
\vspace{8pt}
\noindent
Using the invariant~\ref{cn:ti1}, Lemma~\ref{lm:intersect2}
derives
$|\iknh \cap \ikn| \le |\iknh| - 1$
after which, the inclusion $\iknh \subseteq \tkno
\hunion \Tomb~\histo \hunion \set{z}$ leads to
\[
\tag{\normalsize \arabic{tags}}\refstepcounter{tags}\label{eq:le5}
{\small{
|\iknh \cap \ikn| \le |\tkno \hunion \Tomb~\histo|}}
\]
Combined with~\eqref{eq:le0}, this gives us $\res_1 < \res_2 + 2 \
|\tkno \hunion \Tomb~\histo|$, as shown in Figure~\ref{fig:proof2}'s
postcondition. In words, it asserts that the discrepancy between
$\res.1$ and $\res.2$ is bounded by the size of the tokens, which are
either held by the interfering threads at the end or are spent.
%
%

\begin{figure}[t]
  \centering
\[
{\small{
\!\!\!\!\!\!\!\!
\begin{array}{c}
  ~~~~~~~~\spec{~
  \tkns = \emptyset,
  \hists = \gist_0, \ldots
~} ~\comm{P}
\\[2pt]
  \begin{tabular}{c || c}
$
\spec{\!\!\!
    \begin{array}{c}
    \tkns = \emptyset,
    \hists = \gist_0
  \end{array}\!\!\!}
$
&
$
\spec{\!\!\!\begin{array}{c}
    \tkns = \emptyset,
    \hists = \emptyset
  \end{array}\!\!\!}
$
\\[3pt]
   $\begin{array}{l}
      \res_1 \Asgn \esc{getAndInc();}\\[1pt]
      \res_2 \Asgn \esc{getAndInc();}\\[1pt]
      \kw{return}~(\res_1, \res_2) ~\comm{ =: \res}\!\!\!
    \end{array}$
\!\!\!\!\!\!
& ${\small{e}}$ 
\\\\[-5pt] 
$
\spec{\!\!\!
{\small{
  \begin{array}{c}
    \res.1 < \res.2 + 2 \ |\tkno \hunion \Tomb~\histo|
  \end{array}
}}
  \!\!\!}\!\!$
\!\!
&
\!\!$\spec{\!\!\!
{{
  \begin{array}{c}
    \exists \gist \ldot 
    \hists = \gist, 
    |\gist| = N,
    \ldots
  \end{array}
}}
\!\!\!}$
\end{tabular}
\\\\[-5pt]
\comm{\res_1 := \res.1.1, \res_2 := \res.1.2}
\\\\[-6pt]
\spec{
\res_1 < \res_2 + 2 \ |\tkno \hunion \Tomb~(\histo \hunion \gist)|
}
\\[3pt]
~~~~~~~~\spec{
\res_1 < \res_2 + 2 \ |\tkno \hunion \Tomb~\histo| + 2 \ N
} ~\comm{Q}
\end{array}
}}  
\]
  \caption{Proof outline for the $\eqqc$ program.}
  \label{fig:eqqcproof}
\end{figure}

Figure~\ref{fig:eqqcproof} shows the proof outline for $\eqqc$ via the
spec from Figure~\ref{fig:proof2}.
By the parallel composition rule~\eqref{eq:parrule}, the precondition
splits into two subjective views, where we send the initial history
$\gist_0$ to the left thread, and the empty history to the right
thread. The proof from Figure~\ref{fig:proof2} then applies to the
left thread, and the spec~\eqref{eq:espec} applies to the right
one. Final $\histo$ of the left thread is the union of $\histo$ from
the joined thread with $\gist$, since the environment of the left
thread includes the right thread and of the join. Rewriting by this
property in the postcondition of the left thread gives us the post of
the joint thread: $\res_1 < \res_2 +2\ |\tkno \hunion \Tomb~(\histo
\hunion \gist)|$, which we can next simplify into
\[
\res_1 < \res_2 + 2\ |\tkno \hunion \Tomb~\histo| + 2\ N
\]
because $\Tomb$ distributes over $\hunion$, and $|\Tomb~\gist| =
|\gist| = N$. Finally, we restrict the external interference by
considering $(\hide~\eqqc)$. From the properties of hiding,
we deduce that $\tkno$ and $\histo$ in $Q$ are empty, hence we can
simplify into $\res_1 < \res_2 + 2 \ N$, which is the desired
result~\textbf{R3}.
%

\section{Discussion}
\label{sec:discussion}

%



\paragraph{Reasoning about quantitatively quiescent queues}

The idea of interference-capturing histories, which allowed us to
characterize the out-of-order discrepancies between the results of a
counting network in Section~\ref{sec:counting},
can be applied to specify other balancer-based data structures, for
instance, queues~\cite{Derrick-al:FM14}.
The picture on the right illustrates schematically a non-linearizable
queue~\cite{Derrick-al:FM14}, which is built out of 
\begin{wrapfigure}[6]{r,trim}[-45pt]{1.5cm}   
\centering 
\includegraphics[width=3.0cm]{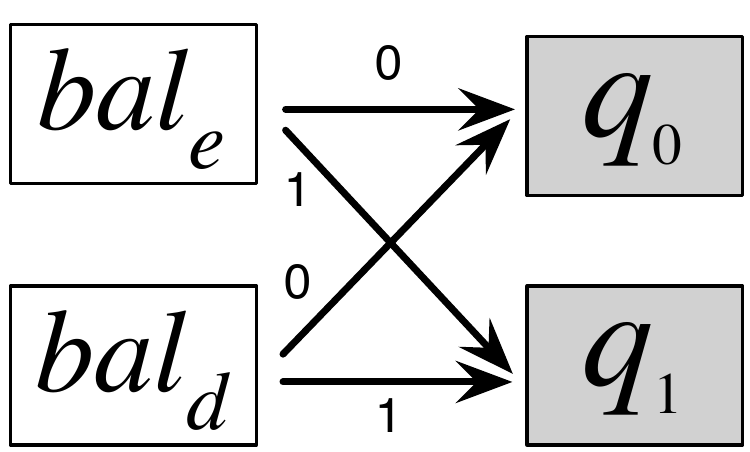} 
\end{wrapfigure}
two \emph{atomic} queues, $q_0$ and $q_1$, and two
balancers, $\bal_e$ and $\bal_d$.  The balancers are used to
distribute the workload between the two queues by directing the
threads willing to enqueue and dequeue elements, correspondingly.

%
One can think of representing the pending enqueue/dequeue requests to
each of the two queues, $q_0$ and $q_1$, by two separate sets of
tokens, as shown in Figure~\ref{fig:chist2}.
The white and gray boxes correspond to the present and
dequeued nodes in the queue in the order they were added/removed.
Therefore, white elements are those that are currently in the queue.
Similarly, the white-colored tokens are for enqueueing elements, so
the elements $x$, $y$, $z$ and $k$ are going to be added to the
corresponding atomic queues. Gray-colored tokens correspond to
dequeueing capabilities for one or another atomic queue, distributed
among the threads, so the elements $c$ and $d$ are going to be removed
next, on the expense of the corresponding dequeue tokens.
The timestamps of the entries in the queue history, omitted from the
figure, are created, as elements are being enqueued to $q_0$ and
$q_1$, and the parity of a timestamp corresponds to the atomic queue
being changed. Thus, there might be ``gaps'' in the combined queue
history reminiscent to the gaps in the counter history from
Section~\ref{sec:counting} (\eg, the gap caused by the absence of an
``even'' element in the combined history right between $d$ and $e$ in
Figure~\ref{fig:chist2}, as indicated by ``?''), which will cause
out-of-order anomalies during concurrent executions.
By accounting for the number of past and present tokens for enqueueing
and dequeueing, one should be able to capture the effects of
interference and express a quantitative boundary on the discrepancy
between the results, coming out of order.

{
\setlength{\belowcaptionskip}{-10pt}
\begin{figure}[t]
\centering
\includegraphics[width=8.2cm]{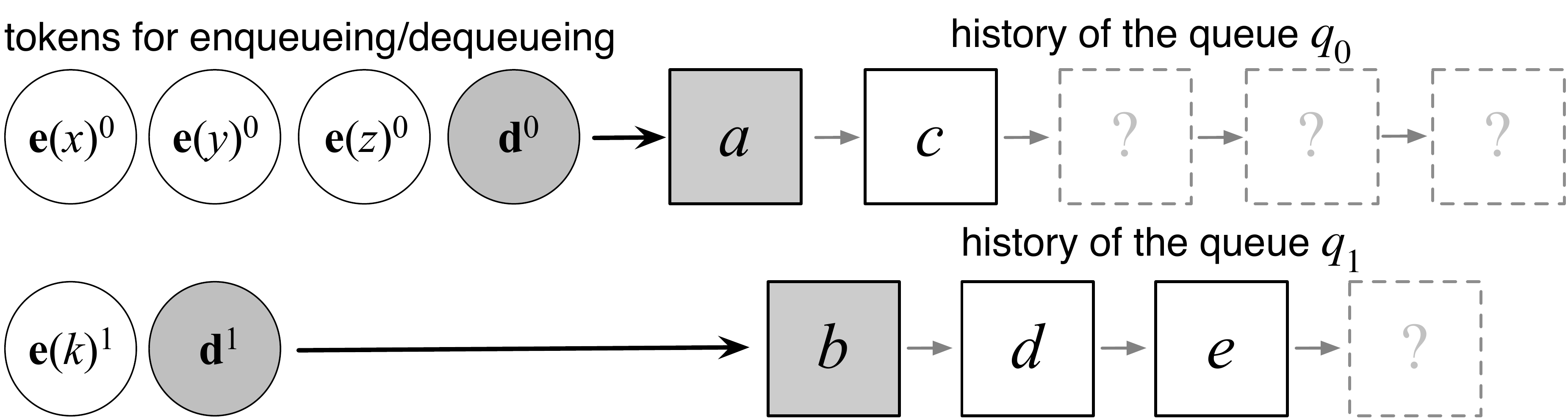}      
\caption{Tokens and histories of a balancer-based queue.}
\label{fig:chist2}
\end{figure}
}





\paragraph{How much information to expose in a spec?}
\label{sec:how-much-information}

The specs we have proved for concurrent objects in
Sections~\ref{sec:overview} and~\ref{sec:counting} allow for efficient
compositional reasoning about clients, but they are also non-trivial
to formulate and verify. Luckily, the FCSL way of reasoning provides a
flexible solution for the compositionality-versus-complexity
conundrum~\cite[\S 7]{Lamport:COMPOS97}.

In FCSL, it is up to the library implementor to decide, how much of
implementation-specific insight should go into a spec. The amount of
such details is determined based on the foreseen client scenarios. For
instance, we have hidden the balancer in the spec~\eqref{eq:qc-spec},
but decided keep the exact constant $2$, which would allow us to
derive more precise quantitative bounds later (see
Section~\ref{sec:qqc-client}). However, we could have hidden this
component too (as well as, for instance, some parts of the invariant
$\ic$), by employing in the specification sigma-types (a
dependently-typed analogue of existential types), provided by FCSL as
it's embedded into Coq~\cite{Coq-manual}. We could have also omitted
tokens from the spec, therefore, reducing the set of derivable
client-specific properties to Section~\ref{sec:counting}'s~\textbf{R1}
only.

\section{Mechanization and Evaluation}
\label{sec:evaluation}

%

In order to assess feasibility of the presented above ideas, we have
mechanized the specs and the proofs of all the examples from this
paper, taking advantage of the fact that FCSL has been recently
implemented as a tool for concurrency
verification~\cite{Sergey-al:PLDI15} on top of the Coq proof
assistant~\cite{Coq-manual}.

Table~\ref{tab:locs} summarizes the statistics with respect to our
mechanization in terms of lines of code and compilation times. 
The examples were proof-checked on a 3.1~GHz Intel Core~i7 OS~X
machine with 16 Gb RAM, using Coq~8.5pl2 and Ssreflect
1.6~\cite{Gonthier-al:TR}.
As the table indicates, a large fraction of the implementation is
dedicated to proofs of preservation of resource
invariants~(\textsf{Inv}), \ie, checking that the actual
implementations do not ``go wrong''.
In our experience, these parts of the development are the most tricky,
as they require library-specific insights to define and reason about
auxiliary histories.
Since FCSL is a general-purpose verification framework, which does not
target any specific class of programs or properties, we had to prove
problem-specific facts, \eg, lemmas about histories of a particular
kind (\textsf{Facts}), and to establish the specs of interest stable
(\textsf{Stab}). Once this infrastructure has been developed, the
proofs of main procedures turned out to be relatively small
(\textsf{Main}).
%
%

{
\begin{table}
{
\sffamily\small 
\centering
\begin{tabular}{|@{\ }l@{\ }||@{\ }c@{\ }|@{\ }c@{\ }|@{\ }c@{\ }|@{\ }c@{\ }|@{\ }c@{\ }||@{\ }r@{\ }|}
  \hline
  \textbf{Program} &  
                     {Facts} & {Inv} &
                                       {Stab} & {Main} & \textbf{Total}
  & \textbf{Build~~~}    
  \\ \hline \hline 
  Exchanger \hfill (\S \ref{sec:exchanger}) & 365 & 1085 & 446 & 162 & 2058 & 4m~~46s
  \\
  Exch. Client \hfill (\S \ref{sec:cal}) & 258 & -- &--& 182 & 440 & 57s
  \\
  Count. Netw. \hfill (\S \ref{sec:counting}) & 379 & 785 & 688 & 27 & 1879 & 12m~23s
  \\
  CN Client 1 \hfill (\S \ref{sec:qc-client}) & 141 &--&--& 180  & 321 & 3m~11s
  \\
  CN Client 2 \hfill (\S \ref{sec:qqc-client})& 115 &--&--& 259 & 374 & 3m~~~9s 
  \\[2pt] \hline
\end{tabular}
}
\caption{
  Mechanization of the examples: lines of code for program-specific facts \intab{Facts},
  resource invariants and transitions \intab{Inv}, 
  stability proofs for desired specs \intab{Stab}, spec and proof sizes for main
  functions \intab{Main}, total LOC count \intab{\textbf{Total}}, and build
  times \intab{\textbf{Build}}. The ``--'' entries indicate the
  components that were not needed for the example.
} 
\label{tab:locs}
\end{table}}

Fortunately, trickiness in libraries is invisible to clients, as FCSL
proofs are compositional. Indeed, because specs are encoded as Coq
types~\cite{Sergey-al:PLDI15}, the substitution principle
automatically applies to programs \emph{and proofs}.
%
%
At the moment, our goal was not to optimize the proof sizes, but to
demonstrate that FCSL as a tool is suitable \emph{off-the-shelf} for
machine-checked verification of properties in the spirit of novel
correctness
conditions~\cite{Hemed-al:DISC15,Aspnes-al:JACM94,Jagadeesan-Riely:ICALP14}.
Therefore, we didn't invest into building advanced
tactics~\cite{McCreight:TPHOL09} for specific classes of
programs~\cite{Zee-al:PLDI08} or
properties~\cite{Dragoi-al:CAV13,Vafeiadis:CAV10,Bouajjani-al:POPL15,Burckhardt-al:PLDI10},
and we leave developing such automation for future work.


\section{Related Work}
\label{sec:related}




\paragraph{Linearizability and history-based criteria.}


The need for correctness criteria alternative to
linearizability~\cite{Herlihy-Wing:TOPLAS90}, which is more relaxed
yet compositional, was recognized in the work on counting
networks~\cite{Aspnes-al:JACM94}.
The suggested notion of quiescent
consistency~\cite{Shavit-Zemah:TOPLAS96} required the operations
separated by a quiescent state to take effect in their logical order.
A more refined correctness condition, \emph{quasi-linearizability},
implementing a relaxed version of linearizability with an upper bound
on nondeterminism, was proposed by Afek~\etal~\cite{Afek-al:OPODIS10},
allowing them to obtain the quantitative boundaries similar to what we
proved in Section~\ref{sec:qqc-client}.
The idea of relaxed linearizability was later used in the work on
\emph{quantitative relaxation} (QR)~\cite{Henzinger-al:POPL13} for
designing scalable concurrent data structures by changing the
specification set of sequential histories.
Most recently, \emph{quantitative quiescent consistency} has been
proposed as another criterion incorporating the possibility to reason
about effects of bounded thread
interference~\cite{Jagadeesan-Riely:ICALP14}.
It is worth noticing that some of these correctness criteria are
incomparable (\eg, QC and QR~\cite{Henzinger-al:POPL13}, QL and
QQC~\cite{Jagadeesan-Riely:ICALP14}) hence, for a particular
concurrent object, choosing one or another criterion should be
justified by the needs of the object's client. Therefore, a suitable
correctness condition is essentially ``\emph{in the eye of the
  beholder}'', as is typical in programming, when designing libraries
and abstract data structures, and the logic-based approach we advocate
provides precisely this flexibility in choosing desired specs.


\paragraph{Hoare-style specifications of concurrent objects.}
\label{sec:related-logic-based}


Hoare-style program logics were used with great success to verify a
number of concurrent data structures and algorithms, which are much
more natural to specify in terms of observable state modifications,
rather than via call/return histories. The examples of such objects
and programs include
barriers~\cite{Dodds-al:POPL11,Hobor-Gherghina:ESOP11}, concurrent
indices~\cite{ArrozPincho-al:OOPSLA11}, flat
combiner~\cite{Turon-al:ICFP13,Sergey-al:ESOP15}, event
handlers~\cite{Svendsen-Birkedal:ESOP14}, shared graph
manipulations~\cite{Raad-al:ESOP15,Sergey-al:PLDI15}, as well as their
multiple client programs.
The observation about a possibility of using program logics as a
correctness criterion, alternative to linearizability, has been made
in some of the prior
works~\cite{Jacobs-Piessens:POPL11,ArrozPincho-al:OOPSLA11,Svendsen-al:ESOP13}.
Their criticism of linearizability addressed its inability to capture
the state-based properties, such as dynamic memory
ownership~\cite{Jacobs-Piessens:POPL11}---something that
linearizability indeed cannot tackle, unless it's
extended~\cite{Gotsman-Yang:CONCUR12}.
However, we are not aware of any prior attempts to capture CAL, QC and
QQC-like properties of concurrent executions by means of \emph{one and
  the same} program logic and employ them in client-side
reasoning.

%

Several logics for proving linearizability or, equivalently,
observational refinement~\cite{Filipovic-al:TCS10,Turon-al:POPL13},
have been proposed
recently~\cite{Turon-al:ICFP13,Liang-Feng:PLDI13,Vafeiadis:PhD}, all
employing variations of the idea of using \emph{specifications as
  resources}, and identifying (possibly, non-fixed or non-local)
linearization points, at which such specification should be ``run''.
In these logics, after establishing linearizability of an operation,
one must still devise its Hoare-style spec, such that the spec is useful for
the clients.

Similarly to the way linearizability allows one to replace a
concurrent operation by an atomic one, several logics have implemented
the notion of \emph{logical atomicity}, allowing the clients of a data
structure to implement application-specific synchronization on top of
the data structure operations.
Logical atomicity can be implemented either by parametrizing specs
with client-specific auxiliary
code~\cite{Jacobs-Piessens:POPL11,Svendsen-al:ESOP13,Svendsen-Birkedal:ESOP14,Jung-al:POPL15}
or by engineering dedicated rules relying on the simulation between
the actual implementation and the ``atomic''
one~\cite{ArrozPincho-al:ECOOP14}.
%

Instead of trying to extend the existing approaches for logical
atomicity to non-linearizable objects (for which the notion of
atomicity is not intuitive), we relied on a general mechanism of
auxiliary state, provided by FCSL~\cite{Nanevski-al:ESOP14}. 
Specifically, we adopted the idea of histories as auxiliary
state~\cite{Sergey-al:ESOP15}, which, however, was previously explored
in the context of FCSL only for specifying linearizable structures.
%
%
%
We introduced enhanced notation for referring directly to histories
(\eg, $\hists$, $\histo$), although FCSL's initial logical
infrastructure and inference rules remained unchanged.

Recently, attempts were made to unify the common idioms occurring in a
number of concurrency logics in a generic framework of
\emph{Views}~\cite{DinsdaleYoung-al:POPL13}.
However, that result is orthogonal to our findings, as \emph{Views}
are a framework for proving logics sound, not to prove programs, and
this paper, we focused on using a particular logic (FCSL) for specifying a
new class of concurrent data structures.

In this work, we do not argue that FCSL is the only logic capable of
encoding custom correctness conditions and their combinations, though,
we are not aware of any other work exploring a similar possibility.
However, we believe that FCSL's explicit \emph{other}
subjective state component provides the most straightforward way to do
so.
The logics like CAP~\cite{DinsdaleYoung-al:ECOOP10} and
TaDA~\cite{ArrozPincho-al:ECOOP14}, from our experience and personal
communication with their authors, may be capable of implementing our
approach at the expense of engineering a much more complicated
structure of capabilities to encode histories and their invariants,
and ``snapshot'' interference of an environment.
Other logics incorporating the generic PCM
structure~\cite{Raad-al:ESOP15,Jung-al:POPL15,Jung-al:ICFP16,Turon-al:OOPSLA14}
might be able to implement our approach, although none of these logics
provide an FCSL-style rule for hiding~\eqref{eq:ehide} as a uniform
mechanism to express explicit quiescence.


Concurrently with this work, Hemed~\etal developed a (not yet
mechanized) verification technique for CAL~\cite{Hemed-al:DISC15},
which they applied to the exchanger and the elimination
stack. Similarly to our proposal, they specify CAL-objects via Hoare
logic, but using one global auxiliary history, rather than subjective
auxiliary state. 
This tailors their system specifically to CAL (without a possibility
to incorporate reasoning about other, non CA-linearizable, concurrent
structures), and to programs with a \emph{fixed} number of threads. In
contrast, FCSL supports dynamic thread creation, and is capable of
uniformly expressing and mechanically verifying several different
criteria, with CAL merely a special case, obtained by a special choice
of PCM. Moreover, in FCSL the criteria combine, as illustrated in
Section~\ref{sec:cal}, where we combined quiescence with CAL via
hiding. Hiding is crucial for verifying clients with explicit
concurrency, but is currently unsupported by Hemed~\etal's method.

\section{Conclusion and Future Work}
\label{sec:conclusion}


We have presented a number of formalization techniques, enabling
specification and verification of highly scalable non-linearizable
concurrent objects and their clients in Hoare-style program logics.
In particular, we have explored several reasoning patterns, all
involving the idea of formulating execution histories as auxiliary
state, capturing the expected concurrent object behavior.
We have discovered that quantitative logic-based reasoning about
concurrent behaviors can be done by storing relevant information about
interference directly into the entries of a logical history.

We believe that our results help to bring the Hoare-style reasoning
into the area of non-linearizable concurrent objects and open a number
of exciting opportunities for the field of mechanized logic-based
concurrency verification.

For instance, in this paper we have deliberately chosen to focus on
simple client programs to showcase the specs we gave to concurrent
libraries. However, any larger program incorporating these examples
can be verified compositionally in FCSL, out of \emph{these clients'
  specs}, via the substitution principles of
FCSL~\cite{Nanevski-al:ESOP14,Sergey-al:PLDI15}, without the need to
deal with concepts such as histories and tokens that are specific to
particular libraries. Given the bounds, which we formally proved in
Section~\ref{sec:qclients}, we believe that the reasoning patterns we
have described will be useful for mechanical verification of larger
weakly-synchronized approximate parallel
computations~\cite{Rinard:RACES}, exploiting the QC and QQC-like
behavior.

Furthermore, by ascribing interference-sensitive quantitative specs in
the spirit of~\eqref{eq:qc-spec} to relaxed concurrent
libraries~\cite{Henzinger-al:POPL13}, one can assess the applicability
of a library implementation for its clients: the clients should
tolerate the anomalies caused by interference, as long as they can
logically infer the desired safety assertions from a library spec,
which is fine-tuned for particular usage scenarios.








\setlength{\bibsep}{1.8pt} 
\bibliographystyle{abbrv}
\softraggedright 
\bibliography{bibmacros,references,proceedings}

\appendix

\section{Exchanger Invariants and Proof Outline}
\label{app:exch}


%

\paragraph{Additional exchanger invariants}

The states in the exchanger state-space must satisfy other invariants
in addition to (\ref{tag:exchanging}). These properties arise from our
description of how the exchanger behaves on decorated state. We
abbreviate with $p \mapsto (x; y)$ the heap
$p \mapsto x \hunion p\!+\!1 \mapsto y$.

\begin{enumerate}[label=(\roman*)]
\item\label{exP} $\heapj$ contains a pointer $g$ and a number of
  offers $p \mapsto (v; x)$, and $g$ points to either $\mathsf{null}$
  or to some offer in $\heapj$.

\item $\hists$, $\histo$ and $\mygather{\pending}$ contain only
  disjoint time-stamps. Similarly, $\perms$ is disjoint from $\permo$.

\item\label{matched} All offers in $\pending$ are matched and owned
  by some thread:
  {\small$\exists t\ldot p \mapsto (t, v, w) \subseteq \pending
    \Leftrightarrow p \in \perms \hunion \permo, p \mapsto (v;
    \Matched w) \subseteq \heapj $}.

\item There is at most one unmatched offer; it is the one linked
  from~$g$. It is owned by someone:
  {\small{
      $p \mapsto (v; \Unmatched) \subseteq \heapj \Longrightarrow p
      \in \perms \hunion \permo, g \mapsto p \subseteq \heapj.  $ }}.

\item Retired offers aren't owned:
  {\small{$p \mapsto (v; \Retired)\!\subseteq\!\heapj\!\Rightarrow
    p\!\notin\!\perms\!\hunion\!\permo$}}.

\item The outstanding offers are included in the joint heap, \ie, if
  $p \in \perms \hunion \permo$ then $p \in \mathsf{dom}\ \heapj$.

\item\label{ex:gapless} The combined history
  $\hists \hunion \histo \hunion \mygather{\pending}$ is gapless: if
  it contains a time-stamp $t$, it also contains all the smaller
  time-stamps (sans 0).

\end{enumerate}

%

\paragraph{Explaining the proof outline}

Figure~\ref{fig:exchanger_proof} presents the proof outline for the
spec~\eqref{tag:exchangespec}.
We start with the precondition, and after allocation in line 2,
$\heaps$ stores the offer $p$ in line 3.

If \code{CAS} at line 4 succeeds, the program ``installs'' the offer;
that is, the state (real and auxiliary) is changed simultaneously to
the modification of $g$. In particular, $p$ is added to $\perms$, and
the offer $p$ changes ownership, to move from $\heaps$ to $\heapj$.
Since $b$ will be bound to $\mathsf{null}$, this leads us to the
assertion in line 7. We explain in Section~\ref{sec:background} how
these kinds of changes to the auxiliary state, which are supposed to
occur simultaneously with some atomic operation (in this case,
\code{CAS}), are specified and verified in FCSL. The assertion in line
7 further states $\mathsf{bounded}\ p\ v\ \gist$. We do not formally
define $\mathsf{bounded}$ here (it is in the proof scripts,
accompanying the paper), but it says that $p$ has been moved to
$\heapj$, \ie, $p \mapsto (v; -) \subseteq \heapj$, and that any
time-stamp $t$ at which another thread may match $p$, and thus place
the entry $p \mapsto (t, v,-)$ into $\pending$, must satisfy
$\mathsf{last}(\gist) < t, \twin{t}$. Intuitively, this property is
valid, and stable under interference, because entries in $\pending$
can be added only by generating fresh time-stamps wrt.~the collective
history $\histo \hunion \mygather{\pending}$, and $\gist$ is a subset
of it.
If \code{CAS} in line 4 fails, then nothing changes, so we move to the
spec in line~15.

At line 8, \code{CAS} succeeds if $x\,{=}\,\Unmatched$, and fails if
$x\,{=}\,\Matched w$. Notice that $x$ cannot be $\Retired$; since we
own $p \in \perms$, no other thread could retire $p$.
If \code{CAS} fails, then the offer has been matched with~$w$. \code{CAS}
simultaneously ``collects'' the offer as follows. By invariant (iii),
and $\mathsf{bounded}\ p\ v\ \gist$, the auxiliary map $\pending$
contains an entry $p \mapsto (t, v, w)$, where $\mathsf{last}(\gist) <
t, \twin{t}$. The auxiliary state is changed to remove $p$ from
$\pending$, and simultaneously place $t \mapsto (v, w)$ into $\hists$.
If \code{CAS} succeeds, the offer was unmatched, and is ``retired'' by
removing $p$ from $\perms$.
Lines 12-13 branch on $x$, selecting either the assertion 10 or 11, so
the postcondition follows.


After reading $cur$ in line 18, by invariant (i), we know that $cur$
either points to $\mathsf{null}$, or to some offer $p \mapsto (w; -)
\subseteq \heapj$.


{
\begin{figure}
\[
{\footnotesize{
\begin{array}{rl}
 \Num{1} & \specK{\{\heaps = \emptyset, \perms = \emptyset, \hists = \emptyset, \gist \subseteq \histo \hunion \mygather{\pending} \}}
\\ 
 \Num{2} & ~~~~ p \Asgn \esc{alloc}~(v, \Unmatched);\\
 \Num{3} & \specK{\{\heaps = p \mapsto (v; \Unmatched), \perms = \emptyset, \hists = \emptyset, \gist \subseteq \histo \hunion \mygather{\pending} \}}\\
 \Num{4} & ~~~~ b \Asgn \esc{CAS}~(g, \esc{null}, p);\\
 \Num{5} & ~~~~ \kw{if}~~b~\esc{==}~\esc{null}~~\kw{then}\\
 \Num{6} & ~~~~ ~~~~ \esc{sleep}~(50);\\
 \Num{7} & \specK{\{\heaps = \emptyset, \perms = \{p\}, \hists = \emptyset, \gist \subseteq \histo \hunion \mygather{\pending}, \esc{bounded}\ p\ v\ \gist \}}\\
 \Num{8} & ~~~~ ~~~~ x \Asgn \esc{CAS}~(p\esc{+}1, \Unmatched, \Retired);\\
 \Num{9} & \specK{\{\heaps = \emptyset, \perms = \emptyset, \gist \subseteq \histo \hunion \mygather{\pending},\hbox{}} \\
\Num{10} & \specK{\hphantom{\}}x = \Matched w \implies \exists t\ldot \hists = t \mapsto (v, w), \mathsf{last}(\gist) < t, \twin t,}\\
\Num{11} & \specK{\hphantom{\}}x = \Unmatched \implies \hists = \emptyset \}}\\
\Num{12} & ~~~~ ~~~~ \kw{if}~~x~~\kw{is}~~\Matched w~~\kw{then}~~\kw{return}~~(\esc{Some}~w)\\
\Num{13} & ~~~~ ~~~~ \kw{else}~~\kw{return}~~\esc{None}\\
\Num{14} & ~~~~ \kw{else}\\
\Num{15} & \specK{\{\heaps = p \mapsto (v; \Unmatched), \perms = \emptyset, \hists = \emptyset, \gist \subseteq \histo \hunion \mygather{\pending}\}}\\
\Num{16} & ~~~~ ~~~~ \esc{dealloc}~p;\\
\Num{17} & \specK{\{\heaps = \emptyset, \perms = \emptyset, \hists = \emptyset, \gist \subseteq \histo \hunion \mygather{\pending}\}}\\
\Num{18} & ~~~~ ~~~~ cur \Asgn \esc{read}~g;\\
\Num{19} & \specK{\{\heaps = \emptyset, \perms = \emptyset, \hists = \emptyset, \gist \subseteq \histo \hunion \mygather{\pending},}\\
\Num{20} & \specK{\hphantom{\}}cur = \esc{null} \vee cur \mapsto (w; -) \subseteq \heapj\}}\\
\Num{21} & ~~~~ ~~~~ \kw{if}~~cur~\esc{==}~\esc{null}~~\kw{then}~~\kw{return}~{\esc{None}}\\
\Num{22} & ~~~~ ~~~~ \kw{else}\\
\Num{23} & \specK{\{\heaps = \emptyset, \perms = \emptyset, \hists = \emptyset, \gist \subseteq \histo \hunion \mygather{\pending}, cur \mapsto (w; -) \subseteq \heapj\}}\\
\Num{24} & ~~~~ ~~~~ ~~~~ x \Asgn \esc{CAS}(cur\esc{+}1, \Unmatched, \Matched v);\\
\Num{25} & \specK{\{\heaps = \emptyset, \perms = \emptyset, \gist \subseteq \histo \hunion \mygather{\pending}, cur \mapsto (w; y) \subseteq \heapj,} \\
\Num{26} & \specK{\hphantom{\}}x = \Unmatched \implies y = \Matched v, \exists t\ldot \hists = t \mapsto (v, w), \mathsf{last}(\gist) < t, \twin t},\\
\Num{27} & \specK{\hphantom{\}}x \neq \Unmatched \implies \hists = \emptyset, y \neq \Unmatched \}}\\
\Num{28} & ~~~~ ~~~~ ~~~~ \esc{CAS}~(g, cur, \esc{null});\\
\Num{29} & \specK{\{\mbox{same as above; the state satisfies (iv) because $y \neq \Unmatched$}\}}\\
\Num{30} & ~~~~ ~~~~ ~~~~ \kw{if}~~x~\esc{==}~\Unmatched~~\kw{then}~~w\Asgn \esc{read}~cur;\kw{return}~(\esc{Some}\ w)\\
\Num{31} & \specK{\{ \heaps = \emptyset, \perms = \emptyset, \gist \subseteq \histo \hunion \mygather{\pending}, \res=\esc{Some}\ w}, \\
\Num{32} & \specK{\hphantom{\}}\exists t. \hists = t \mapsto (w, v), \mathsf{last}(\gist) < t, \twin t\}}\\
\Num{33} & ~~~~ ~~~~ ~~~~ \kw{else}~~\kw{return}~\esc{None}\}\\
\Num{34} & \specK{\{\heaps = \emptyset, \perms = \emptyset, \hists = \emptyset, \gist \subseteq \histo \hunion \mygather{\pending}, \res=\esc{None}\}} 
\end{array}
}}
\]
\caption{Proof outline for the exchanger.}
\label{fig:exchanger_proof}
\end{figure} 
}

At line 24, the \code{CAS} succeeds if $x = \Unmatched$ and fails
otherwise. If \code{CAS} succeeded, then it ``matches'' the offer in $cur$;
that is, it writes $\Matched w$ into the hole of $cur$, and changes the
auxiliary state as follows. It takes $t$ to be the smallest unused
time-stamp in the history $\hist = \hists \hunion \histo \hunion
\mygather{\pending}$. Thus $\mathsf{last}(\hist) < t$, and because
$\hist$ has even size by invariant~(\ref{tag:exchanging}), $t$ must be
odd, and hence $t < \twin t = t+1$. The $t \mapsto (v, w)$ is placed
into $\hists$, giving us assertion 26.  To preserve the invariant
(iii), \code{CAS} simultaneously puts the entry $p \mapsto (t, w, v)$ into
$\pending$, for future collection by the thread that introduced offer
$cur$. But, we do not need to reflect this in line 26.
If the \code{CAS} fails, the history $\hists$ remains empty, as no matching is
done. However, the hole $y$ associated with $cur$ cannot be
$\Unmatched$, as then \code{CAS} would have succeded.
Therefore, it is sound in line 28 to ``unlink'' $cur$ from $g$, as the
unlinking will not violate the invariant (iv), which says that an
unmatched offer must be pointed to by $g$.
Finally, lines 30 and 33 select the assertion 26 or 27, and either
way, directly imply the postcondition.


\end{document}